%% file: bass.tex
\documentclass{aa}

\makeatletter\def\cl@chapter{\@elt{theorem}}\makeatother

\usepackage{graphicx}
\usepackage{amsmath}
\usepackage{natbib}
\usepackage[colorlinks=true,linkcolor=blue,citecolor=blue,urlcolor=blue]{hyperref}
\usepackage{booktabs}
\usepackage{orcidlink}
\usepackage{stfloats}

\usepackage[nameinlink,capitalize]{cleveref}

\crefformat{section}{Sect.~#2#1#3}
\crefformat{figure}{Fig.~#2#1#3}
\crefmultiformat{section}{Sect.~#2#1#3}{ and~#2#1#3}{, #2#1#3}{ and~#2#1#3}

\usepackage{txfonts}
\usepackage{xcolor}
\usepackage[normalem]{ulem}
\usepackage{soul}

\usepackage{silence}
\newcommand{\PSF}{\textrm{psf}}
\newcommand{\tang}{_\textrm{t}}
\defcitealias{guinotShapePipeNewShape2022}{G22}
\newcommand{\Axel}{\citetalias{guinotShapePipeNewShape2022}}

\input{macro}

\input R_g_values

\begin{document}

   \title{UNIONS-3500 Weak Lensing: I. A Galaxy Shape Catalogue in the Northern Sky}

\author{%
F. Hervas-Peters\orcidlink{0009-0008-1839-2969}\inst{1,2}\fnmsep\thanks{fhervaspeters@arizona.edu}
\and
S. Guerrini\orcidlink{0009-0004-3655-4870}\inst{3}
\and
M. Kilbinger\orcidlink{0000-0001-9513-7138}\inst{1}
\and
L. Baumont\orcidlink{0000-0002-1518-0150}\inst{4,5,6}
\and
A. Guinot\orcidlink{0000-0002-5068-7918}\inst{7}
\and
C. Daley\orcidlink{0000-0002-3760-2086}\inst{1}
\and
C. Bonini\inst{1,8}
\and
A. Wittje\orcidlink{0000-0002-8173-3438}\inst{9}
\and
C. Murray\orcidlink{0000-0002-4668-1273}\inst{1}
\and
L. W. K. Goh\orcidlink{0000-0002-0104-8132}\inst{10,11}
\and
A. Paradis\inst{1}
\and
A. Tersenov\orcidlink{0009-0007-5348-6701}\inst{12,13,1}
\and
M. J. Hudson\orcidlink{0000-0002-1437-3786}\inst{14,15,16}
\and
L. Van Waerbeke\orcidlink{0000-0002-2637-8728}\inst{17}
\and
H. Hildebrandt\orcidlink{0000-0002-9814-3338}\inst{9}
\and
S. Fabbro\orcidlink{0000-0003-2239-7988}\inst{18,19}
\and
J.-C. Cuillandre\orcidlink{0000-0002-3263-8645}\inst{1}
\and
A. W. McConnachie\orcidlink{0000-0003-4666-6564}\inst{18}
}

\institute{
Université Paris-Saclay, Université Paris Cité, CEA, CNRS, AIM, 91191 Gif-sur-Yvette, France \\
\email{martin.kilbinger@cea.fr}
\and
Department of Astronomy, Steward Observatory, University of Arizona, 933 North Cherry Avenue, Tucson, AZ 85721-0065, USA
\and
Universit\'e Paris Cit\'e, Universit\'e Paris-Saclay, CEA, CNRS, AIM, F-91191, Gif-sur-Yvette, France
\and
Dipartimento di Fisica - Sezione di Astronomia, Università di Trieste, Via Tiepolo 11, 34131 Trieste, Italy
\and
INAF-Osservatorio Astronomico di Trieste, Via G. B. Tiepolo 11, 34143 Trieste, Italy
\and
IFPU, Institute for Fundamental Physics of the Universe, Via Beirut 2, 34151 Trieste, Italy
\and
Department of Physics, McWilliams Center for Cosmology and Astrophysics, Carnegie Mellon University, Pittsburgh, USA
\and
Aix-Marseille Université, CNRS, CNES, LAM, Marseille, France
\and
Ruhr University Bochum, Faculty of Physics and Astronomy, Astronomical Institute (AIRUB), German Centre for Cosmological Lensing, 44780 Bochum, Germany
\and
Institute for Astronomy, University of Edinburgh, Royal Observatory, Blackford Hill, Edinburgh EH9 3HJ, UK
\and
Higgs Centre for Theoretical Physics, School of Physics and Astronomy, The University of Edinburgh, Edinburgh EH9 3FD, UK
\and
University of Crete, Department of Physics, GR-70013 Heraklion, Greece
\and
Institutes of Computer Science and Astrophysics, Foundation for Research and Technology – Hellas (FORTH), N. Plastira 100, Voutes GR-70013 Heraklion, Greece
\and
Department of Physics and Astronomy, University of Waterloo, Waterloo, ON N2L 3G1, Canada
\and
Waterloo Centre for Astrophysics, University of Waterloo, Waterloo, ON N2L 3G1, Canada
\and
Perimeter Institute for Theoretical Physics, Waterloo, ON N2L 2Y5, Canada
\and
Department of Physics and Astronomy, University of British Columbia, Vancouver, BC V6T1Z1, Canada
\and
NRC Herzberg Astronomy and Astrophysics, 5071 West Saanich Road, Victoria, BC 
\and
Department of Computer Science, University of British Columbia, 2366 Main Mall, Vancouver, BC V6T 1Z4, Canada
}
   \date{Received; accepted}

 
  \abstract
   { Weak gravitational lensing has become a widely used effect to characterise the dark-matter distribution on large scales in the Universe by measuring galaxy ellipticities and their statistical correlations.
   }
   {%
   We present the first weak gravitational lensing catalogue for cosmic-shear cosmology
   of the Ultraviolet Near Infrared Optical Northern Survey (UNIONS). We analyse approximately $3\,500$ square degrees of sky
   area in the Northern Hemisphere, observed in the $r$-band by MegaCam on the Canada-France Hawai'i Telescope, achieving a median seeing of 0.7~arcsec. By covering the northern sky, this footprint offers a large overlap with spectroscopy of the Sloan Digital Sky Survey and the Dark Energy Spectroscopic Instrument.
   }
   {%
   Starting from images calibrated for astrometry and photometry, we describe the steps from image processing to catalogue creation. These steps include masking, source detection and selection, star selection, point spread function (PSF) modelling, shape measurement, and calibration. We conduct extensive validation tests, particularly to assess and mitigate the leakage of PSF ellipticity into galaxy shapes. We demonstrate the robustness of the catalogue by investigating correlations between ellipticity and other observational variables as well as structural elements, such as observer-frame image positions and proximity to bright stars. 
   
   }
   {%
   The final galaxy catalogue contains $62$ million galaxies, corresponding to an effective source density of $4.96$ arcmin$^{-2}$. The ellipticity dispersion, commonly referred to as shape noise, is $\sigma_\epsilon=0.27$. 
   }
   {Initiating the first major cosmological analysis by the UNIONS collaboration, this is the first in a series of five papers which cover the various aspects of a robust cosmic shear analysis. Two companion papers discuss the robustness of the catalogue, one through the level of $B$-mode contamination and another by producing and analysing dedicated image simulations for shear calibration, the other two present cosmological results in real and harmonic space.
   }


   \maketitle
%
 
\section{Introduction}


Cosmic shear denotes the coherent distortion of galaxy images by the intervening large-scale structure through the effect of gravitational lensing. Since it is sensitive to the total matter content, this technique is of primary interest to map the matter distribution in the Universe, which is dominated by dark matter. Cosmic shear probes both the evolution of structure and the geometry of the Universe. Over the last two decades, it has become a main probe of cosmological parameters such as the matter density parameter $\Omegam$, the matter density power-spectrum normalisation $\sigma_8$, or the dark-energy equation-of-state parameter $w$. For reviews, see \cite[e.g.,][]{kilbingerCosmologyCosmicShear2015,mandelbaumWeakLensingPrecision2018,pratWeakGravitationalLensing2026}.

While the core purpose of a shear catalogue is to perform a cosmic shear analysis, various other measurements can be performed with such a data set.  These include the measurements of the total mass and density profile of objects such as the dark-matter haloes of groups \citep[e.g.,][]{Li_halomass_2024} and galaxies \citep[e.g.,][]{Huang_galaxy_mass_2020}, the detection of filaments \citep[e.g.,][]{Xia_filaments_2019}, the study of void distributions and profiles \citep[e.g.,][]{Melchior_voids_2013}, intrinsic alignments of galaxies \citep[e.g.,][]{singhIntrinsicAlignmentsSDSSIII2015}, and mass maps \citep[e.g.,][]{Jeffrey_mass_map_2021_DES}, to name some examples. This plethora of applications has made weak-lensing catalogues a particularly rich and sought-after source of cosmological and astrophysical information. 

Galaxy shape measurement algorithms have evolved significantly in the past two decades in response to the biases affecting weak-lensing shear estimation. Broadly, existing approaches fall into two categories: moment-based methods and model-fitting techniques. Moment-based estimators compute weighted integrals of the light distribution involving powers of the spatial coordinates  and therefore avoid strong assumptions about galaxy morphology \citep{KSB_1995,BJ02,HSM_2003}. In contrast, model-fitting approaches assume a parametric description of galaxy structure and optimize parameters against pixel data; the Sérsic profile \citep{Sersic_1963}, either single-component or bulge–disk decomposed, remains the most widely adopted model, forming the basis of algorithms such as \textit{lensfit} \citep{miller_bayesian_2013} used in the Kilo-Degree Survey \citep{giblin_kids-1000_2021}, and \texttt{im3shape} \citep{Zuntz_2013} employed in the DES Y1 analyses \citep{jarvisScienceVerificationWeak2016,Zuntz_2018}. More recently, \texttt{LensMC} proposed an MCMC-based inference framework for \textit{Euclid} \citep{Congedo_2024}. Another popular choice includes a mixture of Gaussians which presents the advantage of fast convolutions and has been used by the Dark Energy Survey \cite{Hogg_2013,Sheldon_ngmix_2014}.

While some works tried to reach minimally biased estimators, the field has progressively shifted toward calibration-driven strategies \citep{Mandelbaum_great_handbook_2014,Mandelbaum_great_2014}, either through realistic simulations or data-driven corrections. Modern surveys increasingly adopt self-calibrating techniques: in the Dark Energy Survey, \texttt{Metacalibration} \citep{huffMetacalibrationDirectSelfCalibration2017,sheldonPracticalWeaklensingShear2017,gattiDarkEnergySurvey2021} and \texttt{Metadetection} \citep{Sheldon_2019,Yamamoto_2025} estimate calibration factors directly from the data by artificially shearing observed images, leveraging the simplicity of Gaussian profile convolutions \citep{Hogg_2013}. The Kilo-Degree Survey recently applied \texttt{Metacalibration} on a moment-based approach \citep{Yoon_metacal_kids_2025}. Following  the logic of the Dark Energy Survey Y3 \citep{gattiDarkEnergySurvey2021}, we fit a Gaussian model \texttt{ngmix} \citep{Sheldon_2015} with \texttt{Metacalibration} in this work. 

This paper is the first in a series of five publications. It presents the UNIONS galaxy catalogue and provides an overview of weak lensing image processing and overall calibration with \textsc{ShapePipe }(\citealp{farrensShapePipeModularWeaklensing2022},\citealp{guinotShapePipeNewShape2022} [hereafter \Axel]). Paper II \citep{Daley_B_mode_2026} validates the catalogue using B-mode tests and derives angular scale cuts on the shear correlation function and power spectrum. Papers III and IV present cosmological constraints in configuration space \citep{melody} and harmonic space \citep{harmony}, respectively. The fifth paper in the series focuses on validation of the shear calibration with image simulations \citep{milli-vanilli}.

UNIONS is a unique dataset for cosmology in many respects: The Northern Hemisphere has been relatively underserved by photometric surveys. In terms of the combination of depth, area, and image quality, UNIONS images are unprecedented: They are deeper than the Sloan Digital Sky Survey \citep{yorkSloanDigitalSky2000} and the Pan-STARRS $3\pi$ survey \citep{chambersPanSTARRS1Surveys2019}, and cover significantly more area than Subaru Strategic Program (SSP) survey with the Hyper Suprime-Cam \citep[HSC;][]{liHyperSuprimeCamYear2023}. Narrow-band optical images from UNIONS will remain state-of-the-art for many years to come.

With observations in the $r$-band having started in 2017, UNIONS is the last of the Stage-III surveys.
Other weak-lensing Stage-III surveys, notably KiDS, DES, and HSC, have recently published results that build on previous iterations and data releases; in part, they correspond to joint analyses
\citep{desandkidscollaborationY3KiDS1000Consistent2023,jeffersonReanalysisStageIIICosmic2025}. The UNIONS results presented here and in the companion papers are completely independent of those other surveys, and correspond to new constraints on cosmological parameters.

The wide-field imager MegaCam \citep{bouladeMegaCamNewCanadaFranceHawaii2003} mounted in the prime focus on the Canada-France Hawai'i Telescope (CFHT) remains, despite its age, one of the best optical cameras in the world. The CFHT has a long history of providing data for weak gravitational lensing measurements, including the first cosmic shear detection \citep{vanwaerbekeDetectionCorrelatedGalaxy2000}. Dedicated cosmic shear surveys were CFHTLS \citep{hoekstraFirstCosmicShear2006}, and, using the same data, CFHTLenS \citep{heymansCFHTLenSCanadaFranceHawaiiTelescope2012}. CFHTLS published cosmic shear results just after the quest of dark energy was placed into stages \citep{albrechtReportDarkEnergy2006}, and can therefore be considered as one of the first Stage-II surveys. UNIONS, as coming out at the dawn of the next-generation surveys such as Euclid, LSST, and Roman, can be seen as the last of the Stage-III surveys.

\begin{table*}[]
    \centering
    \sidecaption
    \begin{tabular}{l|c|c|c|c}
        Survey & Area & $n_\textrm{eff}$ & $\sigma_\epsilon$ & PSF FWHM   \\
        & [deg$^2$] & [arcmin$^{-2}$]  & (weighted) & [arcsec] \\
        \hline
        KiDS-Legacy & 1,347 &  8.94  & 0.29 & 0.7 ($r$)\\
        DES-Y3  &  4,143 & 5.59& 0.261 & 0.98 ($r$) -- 0.85 ($z$) \\
        DES-Y6 &  4,422 & 8.22 &  0.29 & 0.95 ($r$) -- 0.83 ($z$) \\
        HSC-Y3 & 433 &  22.9& 0.236 & 0.59 ($i$) \\ 
        DECADE & 8,768 & 4.49 & 0.258 & 1.08 ($r$) -- 0.95 ($z$)\\ 
        UNIONS-v0  &  1,700  & 6.8 & 0.35 $({\mathrm{raw})}$ & 0.65 ($r$) \\ 
        UNIONS-3500 & 3,648 & 4.96 & 0.27 & 0.71 ($r$) \\ 
    \end{tabular}
    \caption[Recent weak-lensing surveys and their key properties.]{
Recent weak-lensing surveys and their key properties.
Information is taken from \cite{Wright_data_2024,Wright_cosmo_2025} for KiDS legacy,
\cite{Sevilla_Noarbe_des_2021,abbott_dark_2021,Yamamoto_2025} for DES-Y3 \& Y6,
\cite{Li_HSC_2021} for HSC,
\cite{Anbajagane_decades_2025} and private communication for DECADE, and
\Axel\ for UNIONS v0.
}
    \label{tab:survey_recap}
\end{table*}

The first weak-lensing catalogue of UNIONS data was presented in \Axel, which processed $1{,}600$ square degrees of sky area in the $r$-band. Further, processing of a larger area of around $3{,}200$ deg$^2$ was used in \cite{liBlackHoleHalo2024}. The present paper corresponds to the analysis of $3,648$ deg$^2$.

The purpose of this paper is to make the case that UNIONS-3500 is not only a large and competitive weak-lensing data set, but a catalogue that can be used reliably for cosmology given the appropriate mitigation schemes. To do so, in \Cref{sec:data} we first describe the data and processing choices that define the fiducial sample, then in \Cref{sec:shapes} present the shear calibration strategy, and finally in \Cref{seq:PSF} focus on the point spread function (PSF) systematic effect that is most relevant for this catalogue. PSF-related contamination is the dominant additive systematic and therefore receives special attention throughout the paper. Our goal is to show that, after calibration, mitigation, and validation, the remaining systematic contributions are sufficiently controlled for the cosmological analyses presented in the companion papers.

\section{UNIONS data}
\label{sec:data}

The pointing strategy of UNIONS, as described in \cite{gwynUNIONSUltravioletNearInfrared2025} can be traced back to the CFHT $u$-band survey motivated by near-field cosmology. When the need for complementary optical coverage for \textit{Euclid} became more compelling, the targeting was extended according to principles described in \cite{euclidcollaborationEuclidPreparationDR12025}. The target area was selected to be minimally contaminated by the Galactic plane, objects from the Solar System, and zodiacal light. As a result, UNIONS will ultimately cover a large fraction of the darkest extragalactic sky available in the North, with low Galactic extinction. The survey was designed with a strong homogeneity requirement, which was particularly successful in  the $r$-band when considering the spread of the point source depth distribution and both the median and width of the FWHM distribution of point sources. 


\subsection{Catalogue versions}

Three weak-lensing processing campaigns of UNIONS $r$-band data have been
undertaken and published to date. A first catalogue (v0\footnote{Version numbers prior to v1.3 were assigned post-publication, and do not appear explicitly in the original papers.}) covering $1{,}600$
deg$^2$ was created in $2020$, published in \Axel,
and used subsequently for cosmological constraints from peak counts
\citep{aycoberryUNIONSImpactSystematic2023}, and measuring group halo shapes
\citep{robisonShapeDarkMatter2023}.

A second campaign (v1.0 with sub-versions v1.1 and v1.3, see below) using $3{,}200$ deg$^2$ and
thus nearly doubling the previous area was run in 2022. The PSF model was
switched to MCCD \citep{liaudatMultiCCDModellingPoint2021}. The changes compared
to v0 are discussed in \cite{liBlackHoleHalo2024}. Version v1.0 was released
internally to the UNIONS collaboration, but not used in published work. Additional
cuts were applied to create a more robust catalogue, with subsequent version numbers and use as follows. First, version v1.1 was employed for a direct measurement of intrinsic
alignment \citep{Hervas_Peters_IA_2024}. This catalogue includes objects that \textsc{SExtractor} flagged as deblended (\texttt{FLAGS=2}).
Second, with more conservative cuts on object size, version v1.3 was used in a
number of publications: \cite{liBlackHoleHalo2024} measured the black hole to
halo mass relation from weak lensing; \cite{guerriniGalaxyPointSpread2025}
developed and measured PSF systematics for cosmic shear;
\cite{zhangPointSpreadFunction2024} quantified additive biases from PSF
systematics for lensing-density cross-correlations;
\cite{mpethaCosmologyUNIONSWeak2025} derived cosmological constraints from
lensing cluster profiles, whereas \cite{chengUnionsUNIONSUsing2025} used weak
lensing to probe the dark-matter environment of merger galaxies;
\cite{martinLensingMassMatter2025} quantified void properties from lensing by underdense regions; finally, in \cite{Ahad_clusters_Unions_2025} the lensing profile of clusters was used to classify their dynamical state.

\begin{figure}
\includegraphics[width=1.\linewidth]{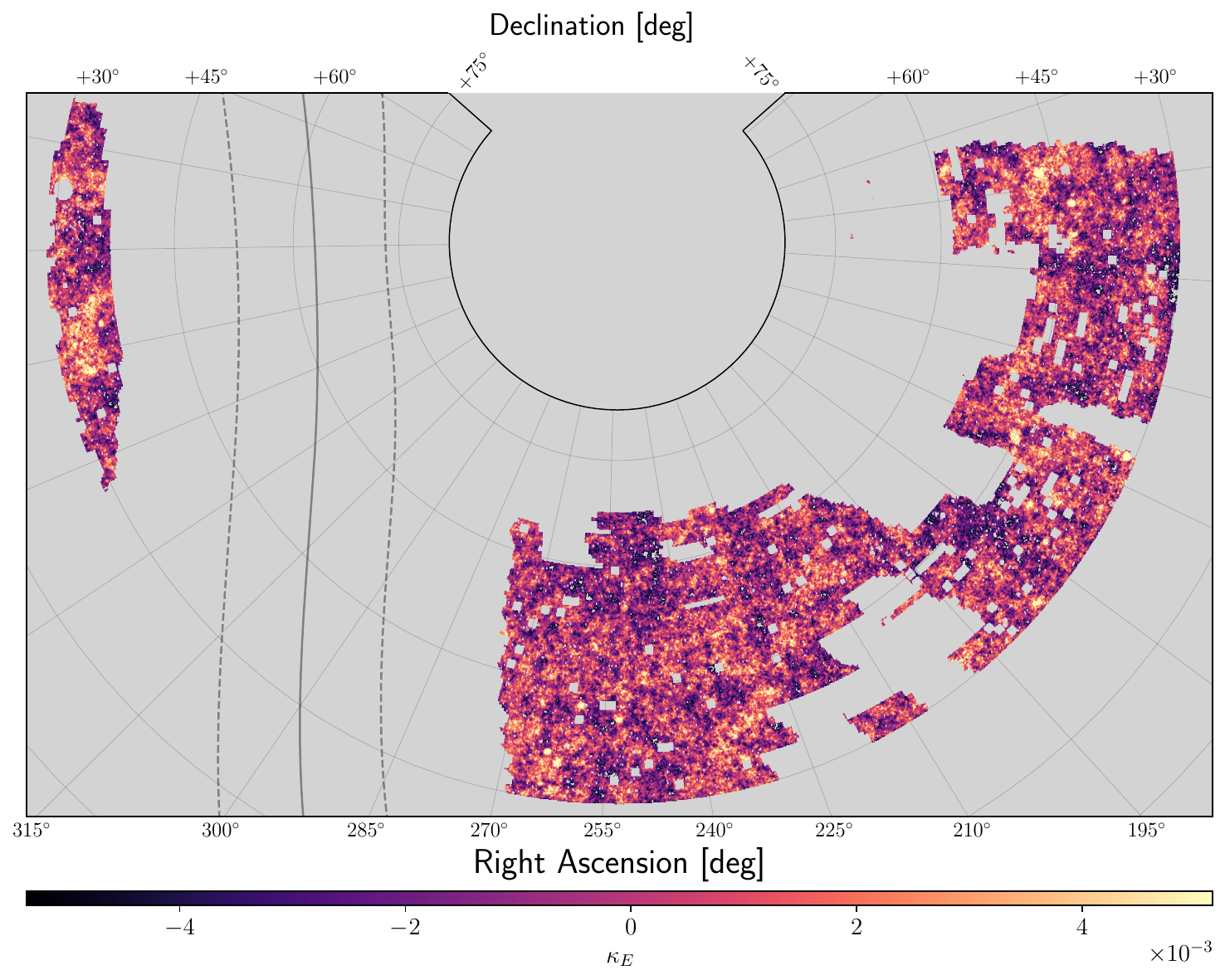}
\caption[Map of the convergence]{Map of the convergence $\kappa$ from the $3{,}500$ deg$^2$ produced with the MCALens algorithm \citep{starckWeaklensingMassReconstruction2021}.}
\label{fig:convergence}
\end{figure}

A re-processing campaign was run in 2023-2024 to reduce PSF leakage. A number of previously
failed tiles were added, bringing the area up to $3\,648$ deg$^2$. The PSF model was
switched back to \textsc{PSFEx} \citep{bertinAutomatedMorphometrySExtractor2011}. This
catalogue is used in the present work and in the companion papers (see \cref{append:cat_versions} for a detailed breakdown). The difference between \textsc{PSFEx} and \textsc{MCCD} will be presented in a dedicated study. To showcase the footprint of UNIONS data used in this work, we present the weak-lensing convergence map in Fig.~\ref{fig:convergence}. It is based on the MCALens algorithm \citep{starckWeaklensingMassReconstruction2021}. The algorithm reconstructs the convergence using Wiener filtering combined with sparse recovery of the non-Gaussian features of the field.

\subsection{Data processing}

\subsubsection{Observation strategy and pre-processing}

The UNIONS $r$-band survey has, on average, three exposures per pixel. This is achieved using single-exposure images with a large dither of a third of a degree, or a third of the focal plane. The exposure time is of the order
200 seconds, while allowing for variations with observing conditions and image quality
(seeing) of $\pm 100$ s, such that the overall depth 
is uniform. The $r$-band utilizes the signal-to-noise ratio queued service observations (SNR QSO) dynamic integration mode, which adapts the exposure time in real time according to observing conditions such as atmospheric absorption, image quality, and sky background. This approach ensures that the combined depth from the three exposures remains uniform across the entire survey footprint.
To detrend the images, night sky flats taken once per dark time run are used for the $r$ band, taken directly from the science images. The bias is computed once a semester. The astrometric calibration is performed by matching to a Gaia EDR3 reference catalogue corrected for proper motion. Only reliable objects (5 $\sigma$ detections in 5 contiguous pixels) are used, leaving typically 20 to 50 objects per CCD. The calibration is estimated to reach 20 mas.
For the photometric calibration an external dataset is used with Pan-STARRS for the $r$-band\footnote{\nolinkurl{https://www.cadc-ccda.hia-iha.nrc-cnrc.gc.ca/en/megapipe/docs/filt.html}}. The zero-point variation is measured for each run, showing some variation over time and appearing to have radial structure across the focal plane. This variation is then removed from the data reaching a 1 mmag precision for most runs. The Pan-STARRS data covers the whole UNIONS footprint and is therefore well-suited for absolute photometric calibration.

The UNIONS exposures are co-added using
\texttt{SWARP} \citep{bertinSWarpResamplingCoadding2010}, to form \emph{tiles} of size $0.5 \times 0.5$ deg$^2$ which have 3\% overlap. They are spaced apart by 0.5 degrees in declination and $0.5/\cos(\mathrm{Dec})$ in right ascension. These tiles are used only for the detection step. The survey reaches a point source depth of $r = 24.1$ at SNR $\nu = 10$ in 2'' diameter apertures. 

\subsubsection{Source detection}

To identify sources, \textsc{ShapePipe} uses the widely adopted \texttt{SExtractor} library \citep{Bertin_1996}. The detections were carried out at the tile level. All detected objects are subject to shape measurement, even though, due to various selections and cuts, only a fraction of objects will make it into the final galaxy catalogue. This fraction as a function of magnitude is shown in Fig.~\ref{fig:cuts_histo}. Our detection process is very conservative and exhaustive, which minimizes detection biases, i.e., spurious correlations between shear imprinted on an object and the probability of the object being detected. We use \texttt{SExtractor}'s flags and masks for a first blend mitigation and deblending step. We then propagate the segmentation map of objects detected by \texttt{SExtractor} to remove neighbouring objects during the shape measurement process. The \texttt{SExtractor} parameters for deblending are shown in Table \ref{tab:sex}; they have been lowered compared to \Axel, to reduce detection biases in the final shear catalogue.

\begin{table}[h]
\centering
\begin{tabular}{|l|l|}
\hline
\textbf{Parameter}       & \textbf{Value}          \\
\hline
\texttt{THRESH\_TYPE}          & RELATIVE                \\
\texttt{DETECT\_THRESH}          & 1.5                     \\
\texttt{DETECT\_MINAREA }          & 5                       \\
\texttt{FILTER}         & Y                       \\
\texttt{FILTER\_NAME}        & default.conv            \\
\texttt{DEBLEND\_NTHRESH }       & 32                      \\
\texttt{DEBLEND\_MINCONT }      & 0.0005                  \\
\hline
\end{tabular}
\caption{\texttt{SExtractor} parametrisation. All other parameters are kept to their default values, following \Axel.}
\label{tab:sex}
\end{table}

\subsubsection{Masking}\label{sec:masking}

We apply different types of masking to define the final shear catalogue.
Mask information is created before, during, and after 
\textsc{ShapePipe} processing. The first, pre-processing masks are generated during MegaCam
processing \citep{Marmo_weight_2008} to flag chip defects and a first, simple pass to detect cosmic rays.

The second set of masks is generated by \textsc{ShapePipe}.
The first mask in this set is generated by the \texttt{mask\_runner} module
of \textsc{ShapePipe}, and flags bright stars, Messier and NGC objects, stellar diffraction spikes,
and stellar reflection halos. The parameters of these masks are based on the \texttt{THELI} pipeline; to mask bright stars we use the Guide Star Catalogue \cite[GSC, version 2.2;][]{VizieR_2001}. We mask stellar halos (diffraction spikes) for stars brighter than magnitude $13$ ($18$), where we use the mean magnitude over all bands available in the GSC. This mask is propagated through \texttt{SExtractor} as $\mbox{IMAFLAGS\_ISO} = 0$.

The second mask in this set is the \texttt{SExtractor} internal flag, indicating basic warnings about the source extraction process. We apply the strict criterion $\mbox{FLAGS} = 0$. This is a conservative choice with consequences on the multiplicative bias discussed in \cref{sec:m_bias}, in particular it removes objects with \texttt{FLAGS} 1 \& 2 that are, respectively, likely to have their flux biased by neighbouring sources by more than 10\% and deblended. 

Next during \textsc{ShapePipe} processing, we apply flags generated by the \texttt{ngmix} shape measurement process. These flags indicate a failed moment calculation as the initial guess for the model fit, or an invalid PSF model at the galaxy position.

Additionally, we remove duplicate objects in the border regions of tiles. These duplicate
objects are due to the small overlapping area between neighbouring tiles.

The third kind of masks is applied in post-processing. These are both pixel-based
and area-based masks. The pixel-based masks are generated by the
neural-network-based software \texttt{maximask} \citep{Maximask_2020} that, very efficiently, removes various contaminants such as cosmic
rays, a residual fringe pattern, diffraction spikes and saturated pixels, among others.

After processing, we noticed that the \textsc{ShapePipe} masks were not large enough to mask some NGC objects entirely, resulting in many fake detections in the outskirts of those objects.
We therefore applied additional pixel-based masks from the \texttt{THELI} pipeline
to conservatively mask bright stars, Messier, and NGC objects.
We also identify regions around stars, split into two stellar magnitude bins, that show faint halo-like emission. These regions can, however, safely be used for most purposes: Shapes of galaxies falling in these regions are in general well measured, and we do not apply this very conservative mask for our shear catalogue, following \cite{Erben_cfhtlens_2013}. We nevertheless tested the impact of excluding these objects and found no significant effects.

An additional area-based mask concerns the
overall exposure coverage: A rough exposure count map is created, and only areas
with $n_\textrm{pointing} \ge 3$ are kept.

Finally, a tile-based flag
indicates whether the tile has been observed and processed in the $r$ band.
This flag is currently redundant, but will be used with corresponding flags for
the $u$, $g$, $i$, and $z$ bands for upcoming multi-band shear tomography catalogues.

The pixel-based masks are transformed into \texttt{healsparse} area maps, and
then the \texttt{healsparse} pixel values are attributed to each catalogue object.

Galaxy-property cuts (size, SNR) do not enter the footprint definition; only spatially defined criteria do (tile coverage, image quality flags, and star and artifact masks), reducing the full $3{,}648~\mathrm{deg}^2$ imaging area to an effective footprint. We compute this footprint by counting \texttt{HEALPix} pixels at $\mathrm{NSIDE} = 4096$ occupied by at least one source in the detection catalogue before galaxy-property cuts, giving $A_{\mathrm{eff}} = N_{\mathrm{pix}} \times 0.74~\mathrm{arcmin}^2$, where $N_{\mathrm{pix}}$ is the number of occupied \texttt{HEALPix} pixels. The footprint covers $2{,}894~\mathrm{deg}^2$. 

\subsubsection{Galaxy selection}

The following criteria define the galaxy sample of the shear catalogue.

\begin{itemize}
  \item $n_{\textrm{ep}} \ge 2$, where $n_{\textrm{ep}}$ is the number of exposures for the given galaxy.
  \item $15 \le r \le 30$, where the $r$-band magnitude is measured via \texttt{SExtractor};
  \item $10 \le \nu \le 500$, where $\nu$ is the signal-to-noise ratio, defined as
  the ratio of flux and flux error as measured during the \texttt{ngmix} modelling. 
  \item $0.707 \le r_\textrm{h} / r_\textrm{h, psf} \le 3$, with $r_\textrm{h}$ being the galaxy half-light radius of the original
  (deconvolved) galaxy image, and $r_\textrm{h, psf}$ the PSF half-light radius;
\end{itemize}

\begin{figure}
\centering
\includegraphics[width=0.9\linewidth]{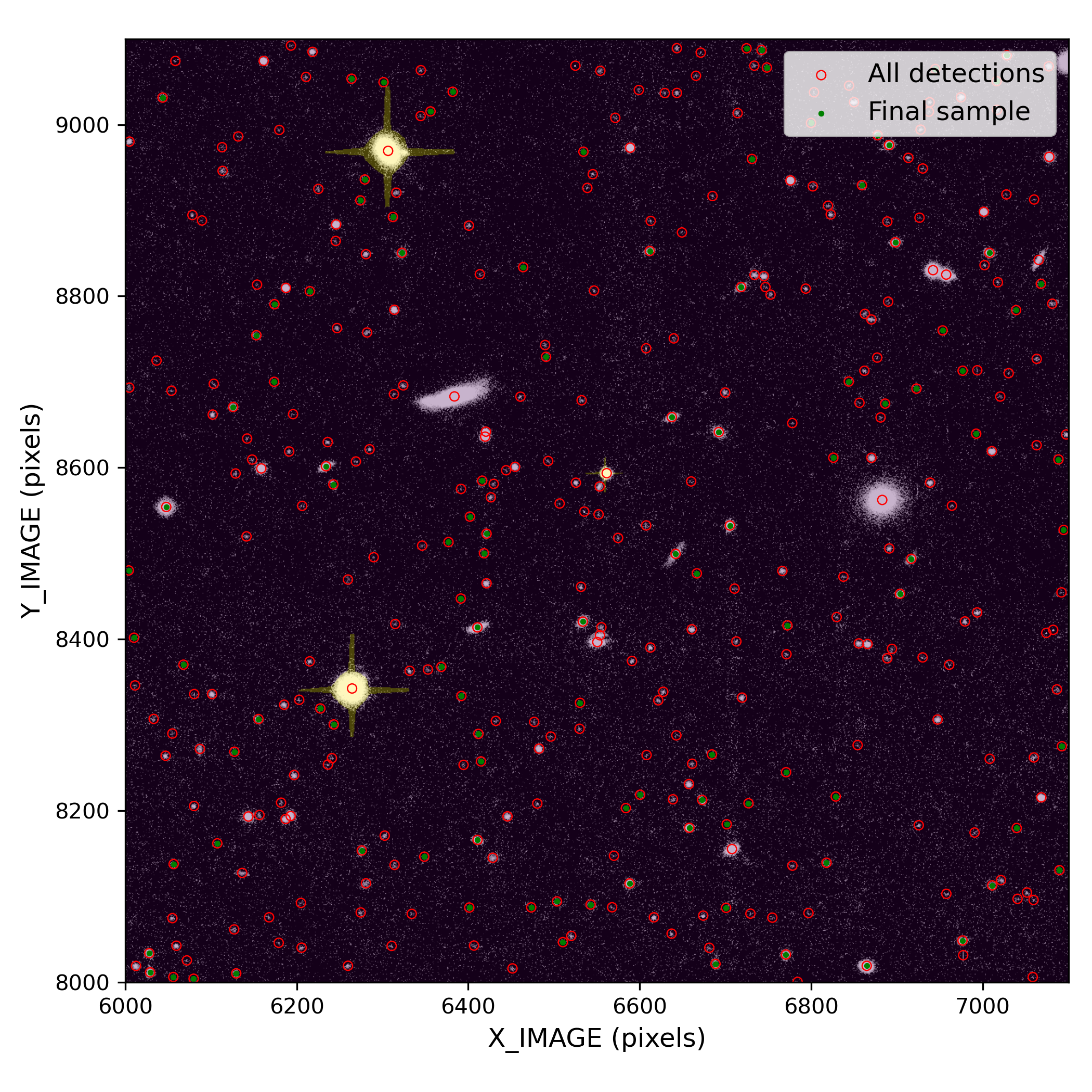}
    \caption{Tile cutout with detected objects and signal sample selection. The yellow areas correspond to the \textsc{ShapePipe} mask. One can see the saturation spikes of bright stars.}
    \label{fig:det_image}
\end{figure}

The latter two selection criteria are applied during \texttt{Metacalibration}, and thus
determine the selection response matrix. An illustration of the detection and selection process is shown in Fig.~\ref{fig:det_image} through a tile cutout. 
Note that we cut on the half-light radius $r_\textrm{h}$, defined before shearing of the object. To relate $r_\textrm{h}$ 
to the Dark Energy Survey size definition, which is based on the galaxy area, one can use the relation 
\begin{equation}
T= \frac{r_\textrm{h}^2}{\ln 2}\frac{1+g_1^2+g_2^2}{1-g_1^2-g_2^2}.  
\end{equation}
Here $g_1$ and $g_2$ are the measured reduced shears. We verified the correspondence of this relation using overlapping objects between the DES-Y3 catalogue and UNIONS.

Note that the $n_\textrm{pointing} \ge 3$ area mask discussed in \cref{sec:masking} is
complementary to the object-based $n_{\textrm{ep}} \ge 2$ selection. The former removes entire tiles that suffer from a low mean exposure, creating a more homogeneous coverage than with the object-based selection, which creates holes of CCD-size and smaller.

\subsubsection{Weights}

As a first estimate, we use inverse-variance weights which depend on the properties of individual galaxies and are presented in \Axel. A shear estimator as the mean over the shapes of galaxies (or any linear combination thereof) has a minimum variance by using inverse variance weights. 

\begin{figure*}
    \centering
    \includegraphics[width=0.55\linewidth]{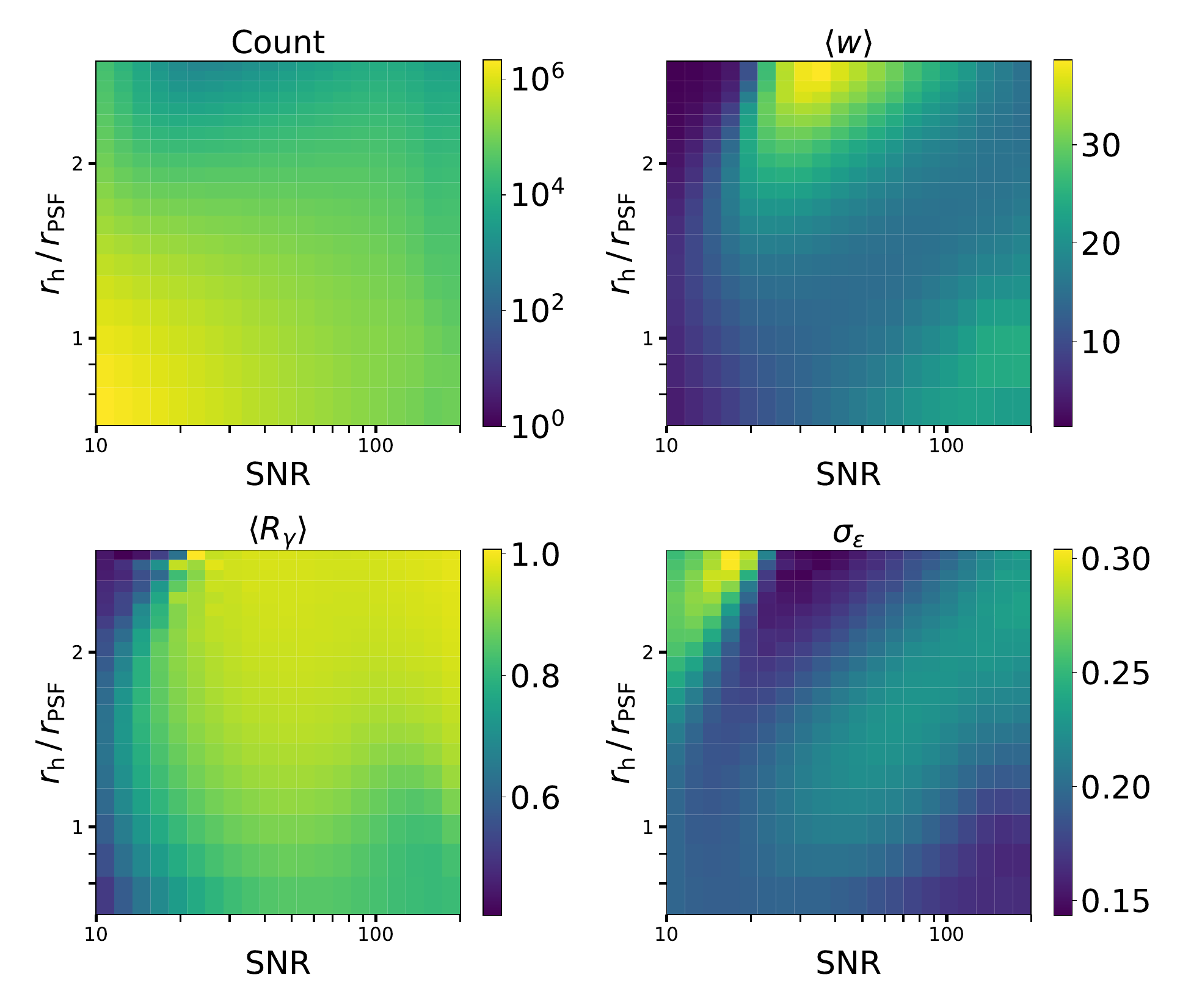}
    \caption{Raw averages of galaxies in bins of ratios of galaxy to PSF size ($y$-axis) and galaxy SNRs ($x$-axis). The binned quantities, indicated by the colour bars, are: Number counts (\emph{upper left}), weights (\emph{upper right}), average diagonal shear response matrix (\emph{lower left}), and average per-component shape noise (\emph{lower right}).
    }
    \label{fig:2d_histograms}
\end{figure*}

We further compute weights that are less sensitive to noise in the estimated measurement errors. Following \cite{gattiDarkEnergySurvey2021}, we compute the weights as piecewise constant values in bins of relative size and SNR. These two quantities are informative predictors of galaxy weight, and we set
\begin{align}\label{equ:weights}
    w\left( \frac{r_{\mathrm{h}}}{r_{\mathrm{psf}}}, \nu \right)
    = \sigma_\gamma^{-2} \left( \frac{r_{\mathrm{h}}}{r_{\mathrm{psf}}}, \nu \right)
    = \left[ \sigma_\epsilon \left\langle R_\gamma\right\rangle \right]^{-2}
    \left( \frac{r_{\mathrm{h}}}{r_{\mathrm{psf}}}, \nu \right) .
\end{align}
The shear dispersion $\sigma_\gamma$ can be written as the product of the ellipticity dispersion $\sigma_\epsilon$ and the shear response. 
The motivation to introduce these weights was to decouple weights from ellipticities, since the shape measurement error in the simple inverse variance weights can correlate with shear. The distribution of these weights as a function of SNR and size-ratio is shown in Fig.~\ref{fig:2d_histograms}. Most galaxies lie in the small-size, low-SNR regime, where weights are small. The weight distribution shows a complex pattern. In particular the top left corner could be affected by blended galaxies, i.e. large diffuse objects, which seem to be strongly down-weighted.

\subsubsection{The final tile sample}

Of the $14{,}596$ tiles, $9$ ($0.06\%$) failed to finish processing, for the following reasons. Two tiles had zero weight everywhere, due to some quirk in the CFHT pipeline. Three tiles had an exceptionally large number ($\sim 10$ times the mean) of \texttt{SExtractor} detections due to their vicinity to a nearby Messier galaxy. Since all those detections fall into subsequently masked regions, we removed those tiles from further processing. Finally, processing of four tiles failed after multiple re-tries for unknown reasons.

\section{Shapes, shears, and calibration}
\label{sec:shapes}

\subsection{Number density}
\begin{figure*}
  \begin{center}
    \includegraphics[trim={3cm 0 3cm 0},clip,width=0.44\linewidth]{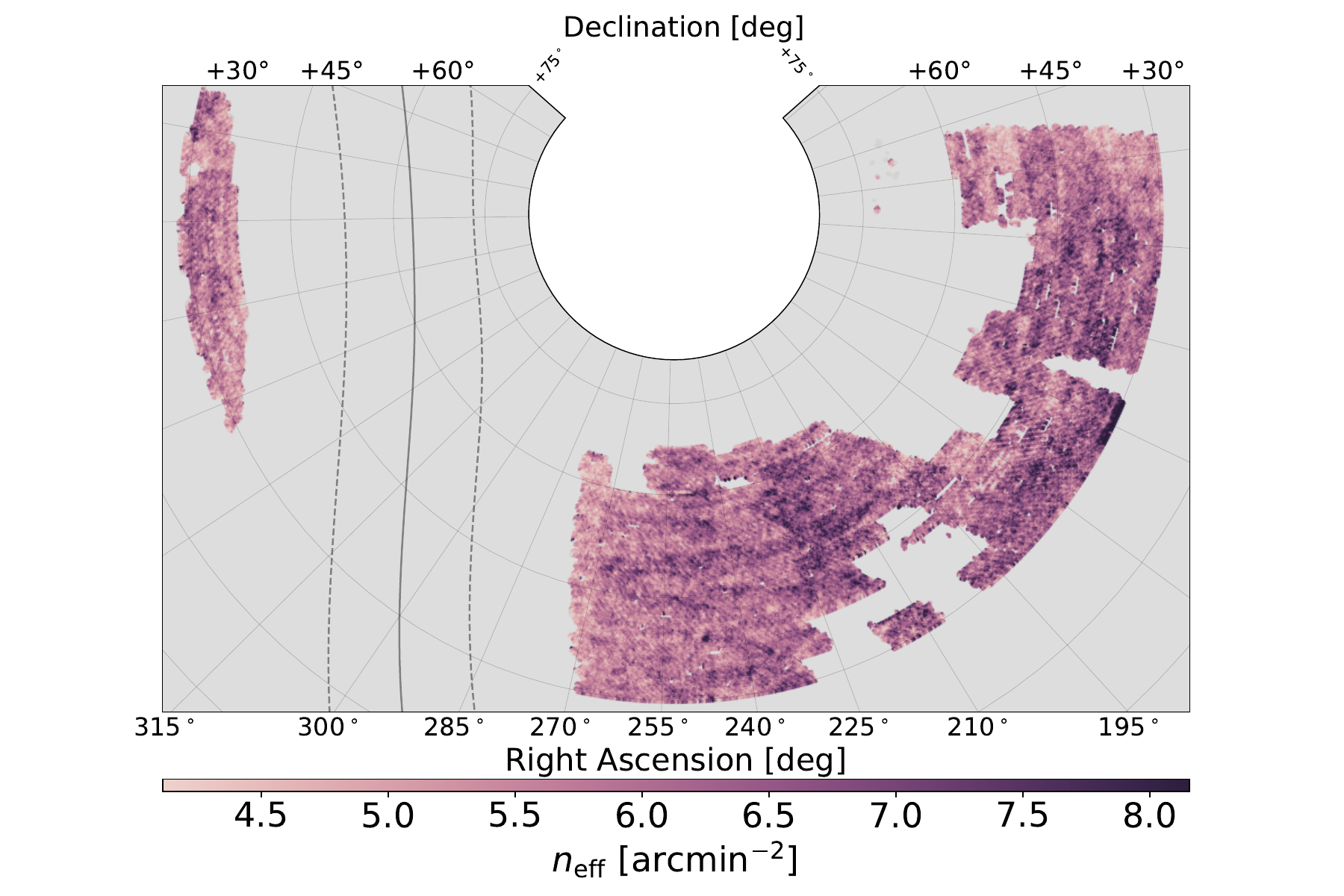}
     \includegraphics[trim={3cm 0 3cm 0},clip,width=0.44\linewidth]{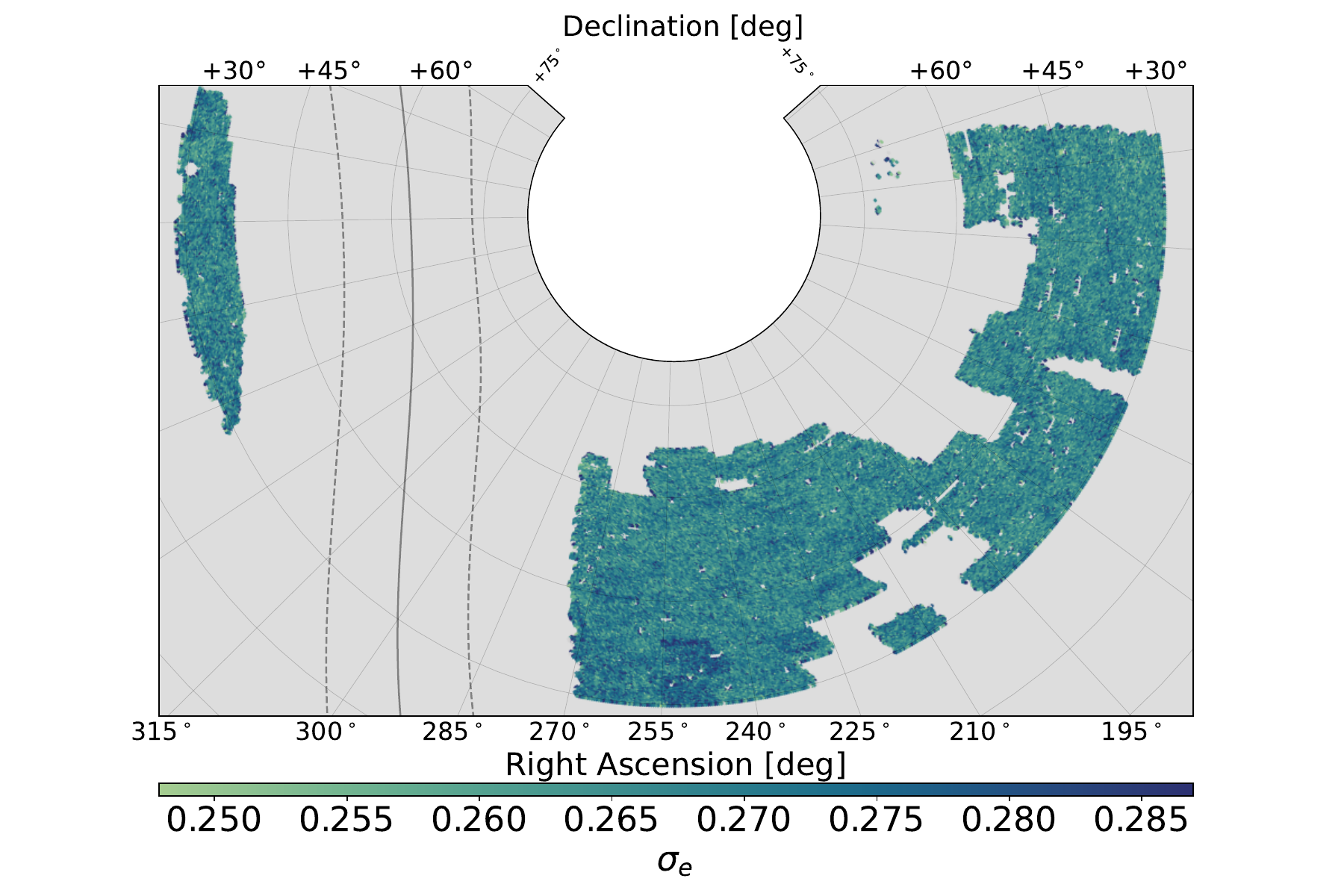}
  \end{center}

  \caption{Footprint of the UNIONS-3500 catalogue showing the effective number density $n_{\mathrm{eff}}$ defined in \cref{eq:neff} and average per-component shape noise $\sigma_\epsilon$ binned with \texttt{HEALPix} pixels NSIDE=512. The density likely decreases towards the Galactic plane due to the higher dust extinction and diffuse galactic light. The distribution of $\sigma_\epsilon$ appears very homogeneous across the footprint.}

  \label{fig:n_eff_sigma_maps}
\end{figure*}

We compute the number density using the definition of \citet{heymansCFHTLenSCanadaFranceHawaiiTelescope2012},
\begin{equation}\label{eq:neff}
    n_{\mathrm{eff}}=\frac 1 A \frac{\left(\sum_{i} w_i\right)^2}{\sum_{i} w^2_i},
\end{equation}
where $w_i$ is the shear weight per galaxy and $A$ the area of the survey. In \cref{fig:n_eff_sigma_maps} the distribution of the effective number density is shown across the survey footprint. A Gaussian smoothing was applied with $\sigma=$ 9 arcmin. The masked area has been subtracted from each pixel. This is important because the raw number density correlates strongly with right ascension due to the stellar density increasing towards the galactic plane as shown in \cref{fig:star_density_residuals}. The spread of $n_{\mathrm{eff}}$ seems to be well constrained. This can be attributed to the homogeneous data in the $r$-band shown in \cite{gwynUNIONSUltravioletNearInfrared2025} and \cref{fig:mag_fwhm_psf_map}. A small decrease towards the galactic plane can hypothetically be attributed to dust extinction.

\subsection{Additive bias}

\begin{table}
\caption{Additive bias components $c_1$ and $c_2$, for different weightings. Central value and error are jackknife estimates.}
\begin{tabular}{lcc}
\toprule
version & $c_1 / 10^{-4}$ & $c_2 / 10^{-4}$ \\
\midrule
DES weights & $-1.6 \pm 0.3$ & $2.3 \pm 0.3$ \\
inverse-variance & $-1.4 \pm 0.3$ & $2.3 \pm 0.3$ \\
no weights & $-1.1 \pm 0.3$ & $2.4 \pm 0.3$ \\
\bottomrule
\end{tabular}
\label{tab:c}
\end{table}

When measuring gravitational shear, a common parametrisation to relate observed and true shear is  
\begin{equation}
    \gamma_{\mathrm{obs}}=(1+m)\,\gamma_{\mathrm{true}}+c\ ,
\end{equation}
where $m$ and $c$ are called multiplicative and additive biases respectively. \Cref{tab:c} shows the mean and associated errors of the additive bias $c_1$ and $c_2$. We remove these values from each ellipticity component to empirically account for the additive bias. These values are relatively small and well aligned with other Stage-III surveys. In DES Y3 (DES Y6) the reported values are $c_1=3.5\times10^{-4} \ (1.9\times10^{-4})$ and $c_2=0.6\times10^{-4} \ (-0.3\times10^{-4})$. This places our average shear at the same order of magnitude. As the error bars indicate this is nevertheless significantly non-zero, indicating an unknown source of bias. During our work on PSF contamination mitigation subtraction, no indication appeared that this global calibration was in conflict with the leakage correction.

The uncertainty on the additive bias using a jackknife estimate is $3 \times 10^{-5}$, which is a factor of $4$ smaller than the requirement on the uncertainty of $c$ for \textit{Euclid} requirements \citep{Cropper_bias_2013}. 

\subsection{Metacalibration shear and selection response}\label{sec:metacal}

As mentioned in the introduction, multiple shape measurement methods have been explored by the weak lensing community in the past decade. Some rely on estimating moments of the surface brightness profiles, others try fitting a parametric profile to the galaxies, such as a mixture of Gaussians or a bulge+disk model using Sérsic profiles. While each has its own advantages and biases, a method widely employed in recent years to estimate model, noise, and selection biases is \texttt{Metacalibration}. This technique quantifies the capacity of a given shape measurement method to recover a known input shear $\Delta \gamma$ that is applied to the data directly \citep{huffMetacalibrationDirectSelfCalibration2017,sheldonPracticalWeaklensingShear2017}. This is done by first deconvolving the image from the PSF model, shearing the deconvolved image and then reconvolving it by a slightly dilated PSF to suppress high frequency modes. During the shearing process, correlated noise can appear, causing biases in the recovered shear. To cancel out the effect of correlated noise, artificial noise is introduced, referred to as \textit{fixnoise}. This artificial noise causes an approximate 20\% loss in SNR. From a theoretical point of view the shear response is obtained by Taylor expanding the ellipticity around a small variation in shear:
\begin{equation}
    e= e|_{\gamma=0}+\left.\frac{\partial e}{\partial \gamma}\right|_{\gamma=0}\ \gamma+ ...
\end{equation}
This leads to defining at first order the shear response matrix $R$ as
    $R=\left.\frac{\partial e}{\partial \gamma}\right|_{\gamma=0} .$
The response matrix is commonly split into a shear response $R_\gamma$ and selection response $R_s$. The shear response can be estimated per galaxy as
\begin{equation}\label{equ:r_gamma}
    R_{\gamma, ij} =\frac{ e_i^{j, +} - e_i^{j, -}}{2 \Delta \gamma},
\end{equation}
where the index $i$ refers to the ellipticity component and the superscripts $j, \pm$ to the branch sheared by $\pm \Delta \gamma$ in component $j$. \texttt{Metacalibration} can calibrate cuts on quantities which are measured by the different branches, by estimating the derivative at the value of the cut via finite differences. The selection responses can only be defined for a global population, since the goal is to quantify how cuts affect the mean ellipticity,
\begin{equation}\label{equ:m_sel}
    R_{\mathrm{sel}, ij}=\frac{\langle e_i^\mathrm{NS}\rangle^{S_{j, +}}-\langle e_i^\mathrm{NS}\rangle^{S_{j, -}}}{2 \Delta \gamma} \ .
\end{equation}
The superscript $_{j,\pm}$ refers to the artificially sheared (in component $j$) galaxy branch, after which the selection cuts are applied. The `NS' indices indicate the ``No Shear'' branch. Calibrated ellipticities are obtained by averaging over shear responses, since individual realisations can be very noisy. We use the full response matrix to calibrate the ellipticities  $\mathbf{e}_{\mathrm{cal}}=\langle{\mathbf{R}}\rangle^{-1}\mathbf{e}_{\mathrm{meas}}$. Weights need to be applied consistently in the numerator and denominator when taking these averages. 
The selection response in our case is $R_{\textrm{sel},1}=0.016$ and  $R_{\textrm{sel},2}=0.015$. While the total shear response is typically below $1$, the selection response can be positive or negative, meaning selection effects can dilute or boost the total shear. \citet{Zuntz_2018} showed that cutting on SNR creates a positive selection response while filtering on size causes a negative one.

In practice \textsc{ShapePipe} uses \texttt{ngmix} to measure shapes with a single Gaussian component. The details of the pipeline are laid out in \Axel. In brief, a guess is made based on a fast moment computation using the HSM algorithm \citep{HSM_2003}. The minimiser uses regularization priors, and a set of six parameters is fitted to each galaxy, [$\Delta x, \,   \Delta y \, , \,g_1, \, g_2, \, r_{50}, \,F$] where $\Delta x$ and $\Delta y$ represent a global offset accounting for a difference between the \texttt{SExtractor} detection and the best fit centre. The ellipticity components are measured as the reduced shears $g_1,g_2$; $r_{50}$ is the half-light radius and $F$ is the flux. All epochs are fitted simultaneously, each with its specific PSF. An issue was discovered during the pipeline validation process, which was the lack of coherent offsetting between the different epochs. The problem was that WCS offsets between pixels with respect to the \texttt{SExtractor} detection were not correctly propagated at the shape measurement stage, meaning that the model was centred on the respective pixel centre of each exposure and not on the common tile detection position. This issue is described in detail in paper V, and its consequences are studied using image simulations.

\begin{figure}
    \centering
    \includegraphics[width=0.95\linewidth]{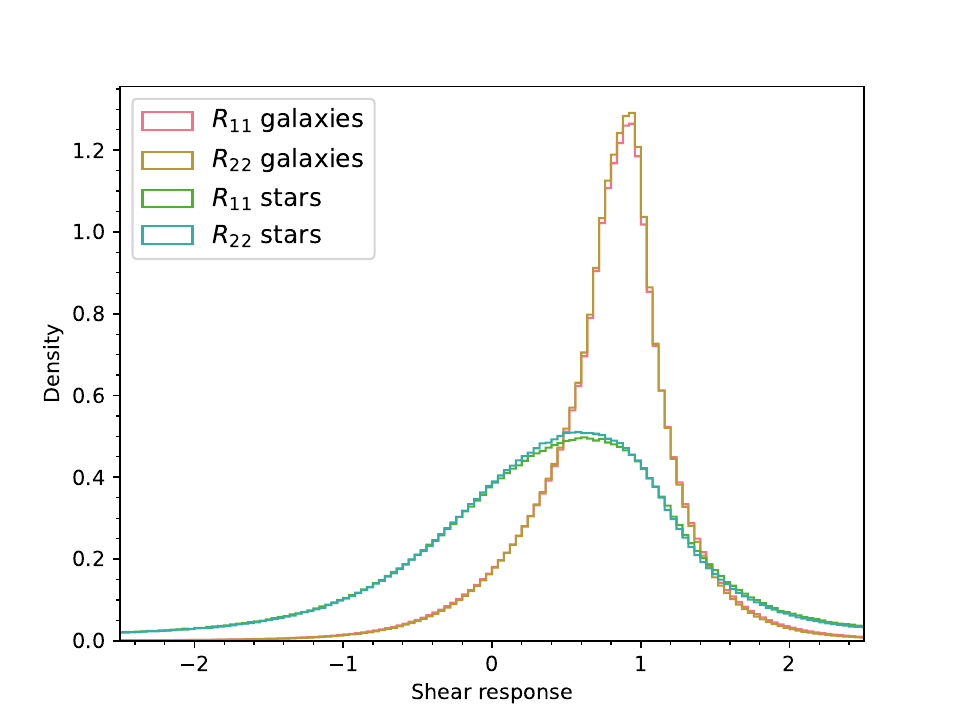}
    \caption{Shear response of stars and galaxies. See text for an explanation of the non-zero shear response in stars.}
    \label{fig:shear_response}
\end{figure}

The shear response distributions in Fig.~\ref{fig:shear_response} show consistency between $R_{11}$ and $R_{22}$ for galaxies. The distribution peaks around unity but has a long tail towards lower values, leading to the mean values $R_{11} = \Rgaa$ and $R_{22} = \Rgbb$. The large non-zero shear response in our star sample can be explained by the offset problem mentioned in the previous paragraph: Stars are point-like objects, and their shear response should average to zero. Paper V describes how this issue was both reproduced and corrected in simulated images by fixing offsetting in the pipeline. This correction will be applied in the upcoming improved UNIONS shape measurements processing.

\subsection{Residual multiplicative bias from image simulations}\label{sec:m_bias}

\subsubsection{Image simulations setup}

In Paper V we describe the production of an extensive set of simulated images. These are built to both validate the shape measurement pipeline and to quantify residual multiplicative biases tied to blends. These synthetic images use a realistic catalogue of galaxies whose distribution is based on an $N$-body simulation in \cite{liKiDSLegacyCalibrationUnifying2023}, and galaxy properties are assigned via a semi-analytic model \citep{Lagos_Shark_2018}. The observation strategy of UNIONS in the $r$-band is reproduced with high fidelity and PSFs are directly drawn from the model estimated on the data.

Galaxies are then simulated in two different configurations, first they are distributed on a lattice for validation and then we use realistic placements from an $N$-body simulation to capture the effects of blended objects. We set galaxies on a grid, since  \cite{sheldonPracticalWeaklensingShear2017,huffMetacalibrationDirectSelfCalibration2017} predict no expected bias when using \texttt{Metacalibration} on isolated galaxies, with a precision on the residual multiplicative bias of $\Delta m \sim \mathcal{O}(10^{-4})$. This serves as a joint validation of our shape measurement pipeline and of the image simulations. It is, of course, possible that different errors conspire to compensate. However, we interface a validated image simulation suite developed by the KiDS collaboration with a well-tested shape measurement code used by DES, giving us confidence that cancelling-out errors are unlikely, in particular at the relatively coarse precision of $(\sigma_m \sim 10^{-2})$ we try to achieve. 

In a second exercise we place galaxies at realistic positions coming from an N-body simulation, and we also include a stellar population generated by \texttt{Trilegal} \citep{Girardi_trilegal_2005} for realistic magnitude distributions. We therefore simulate 40 deg$^2$ of galaxies on a grid spaced by 15 arcseconds and 40 deg$^2$ of galaxies with realistic placements, each configuration being simulated 4 times ($\gamma_1=\pm0.025$, $\gamma_2=\pm0.025$).

\subsubsection{Image simulations results}

The validation exercise revealed two shortcomings of the shape measurement pipeline which require corrective multiplicative bias calibrations. The first effect has already been described above and concerns the lack of WCS offset propagation resulting in non-zero mean shear responses for stars. This issue was identified at the validation stage and a patch in the code allowed us to recover stars that properly deconvolve to point-like objects, indicating a correct propagation of the WCS information. Unfortunately a full re-run was computationally not compatible with the collaboration's objectives due to time constraints. Running a simulation on a grid with and without the offset bug led to a diagnostic of $\Delta m_{\mathrm{offset}}\approx-0.02$. 

The second effect is a selection bias from the \texttt{SExtractor} Flags selection which cannot be accounted for by \texttt{Metacalibration}. The core idea is that, when running full positively and negatively sheared image suites, one does not only verify if the shear values are correctly retrieved, but also if selection effects are properly accounted for, since the same cuts as applied on the data are reproduced on the total simulated catalogue. The selection effect induced by the ellipticity dependence of the \texttt{SExtractor} flags is not captured by the \texttt{Metacalibration} framework  as the flags are constant between the different branches. As stated in \cref{sec:masking} and shown in Fig.~\ref{fig:cuts_histo}, we select galaxies with \texttt{SExtractor} \texttt{FLAGS=0}. The initial motivation for this cut was that objects with \texttt{FLAGS=1 \& 2} might carry residual biases from blends and are therefore not suited for the final shape catalogue. Through the image simulations it was revealed that this selection had very little effect on the amount of blending bias. Due to the low \texttt{DEBLEND\_MINCONT} value  (see Table~\ref{tab:sex}), a large number of objects are flagged as deblended objects, even for galaxies placed on a grid and with their light distribution following a Sérsic profile. This can be understood as noise fluctuations in the outskirts of galaxies producing the necessary intensity contrast to trigger deblending. The multiplicative bias is introduced because more elliptical objects tend to trigger the \texttt{SExtractor} deblending at a higher rate, as their flux is distributed across more pixels. This results in an uncalibrated selection effect of around $\Delta m_{\mathrm{flag-sel}}\approx-0.02$.  
In DES \citep{gattiDarkEnergySurvey2021}, objects were selected with \texttt{FLAGS}$\leq$2.
When we attempted the same selection, a B-mode presence on all scales in configuration and harmonic space made us decide to keep excluding objects with \texttt{FLAGS}$>$0 and apply the multiplicative bias calibration from image simulations. The origin of this B-mode presence is currently not understood. 

In DES, neighbouring objects were treated more comprehensively. They applied a technique called Multi Object Fitting (MOF) \citep{Drlica-Wagner_DES_2017} which relies on a more conservative masking procedure called \textit{\"uberseg} \citep{2016MNRAS.460.2245J}. In short, the mask is built such that all pixels are attributed to the closest object, not only those covered by the \texttt{SExtractor} segmentation map and the fit is then performed iteratively by subtracting the light from neighbours using the models from the previous fit. In our case, we fill the pixels of the \texttt{SExtractor} segmentation map with a Gaussian noise realisation to remove the effect of neighbouring objects. It is unclear if the difference in the masking procedure of neighbouring objects could produce a B-mode correlation. We will investigate this further in coming releases.

The conclusion of the validation efforts, consisting of galaxies placed on a grid, was that our pipeline shows no detectable bias at the precision of our simulations when \texttt{FLAGS}$\leq$2 are included and the WCS offset is correctly propagated during the shape measurement stage.

Finally, since we are not using \texttt{Metadetection}, we quantify the multiplicative bias from blended objects, and find an additional shear bias of $\Delta m_{\mathrm{blend}}\approx-0.017$, consistent with previous surveys \citep{maccrann_dark_2021,li_kids-legacy_2023}. While our statistical precision does not allow us to clearly separate each effect, the total shear bias is $m=-0.0572\pm0.0047$. Since we ran into multiple issues, we conservatively triple our error on the multiplicative bias resulting in $m=-0.0572\pm0.0141$, which we apply at the cosmological inference stage in Papers III and IV. While this calibration factor and its uncertainty are substantially larger than the latest precision of other Stage-III surveys, current efforts for a rerun are ongoing to implement the improvements in an updated \textsc{ShapePipe} version for a complete rerun.

\section{PSF estimation and leakage quantification}
\label{seq:PSF}

In a weak-lensing survey, a precise characterisation of the PSF is crucial to decouple the systematic shape deformations induced by the optical system and atmospheric effects from those that are of cosmological origin. This subject has therefore received dedicated attention in the past decade, both by improvements of the PSF model \citep{jarvis_dark_2021,Schutt_PSF_DESY6_2025} through more complex and complete modelling as well as through more involved diagnostic efforts and mitigation strategies \citep{zhangGeneralFrameworkRemoving2023,Berlfein_roman_psf_2026}.  
In this section we validate the PSF model, assess PSF systematics, quantify the impact of the object-wise leakage correction and estimate the amplitude of the PSF systematics additive bias on the two-point correlation function.

\subsection{Star sample selection and properties}\label{sec:star_sel}
To build a data-driven PSF model, one can rely on a large distribution of point sources across the sky in the form of stars. We identify stars in the size-magnitude space, using the mode of the size distribution per CCD with a $\pm 0.2$ pixel tolerance. Objects with an apparent magnitude $r\in[18,22]$ are kept in the star sample. This procedure is illustrated in Fig.~\ref{fig:mag_rad_plot}. The lower cut excludes objects that can be affected by the brighter-fatter effect or saturation, while the upper bound excludes slightly extended, faint galaxies. Due to the noise of the shape measurement process, some stars end up in the final galaxy sample, potentially biasing the shear measurements, which will be tested in Paper V. The stars are split into two samples: the training sample, on which the PSF model is built, contains 80\%, while the validation sample comprises the remaining 20\% and is used for the diagnostic and calibration work below.

\begin{figure}
    \centering
    \includegraphics[width=\linewidth]{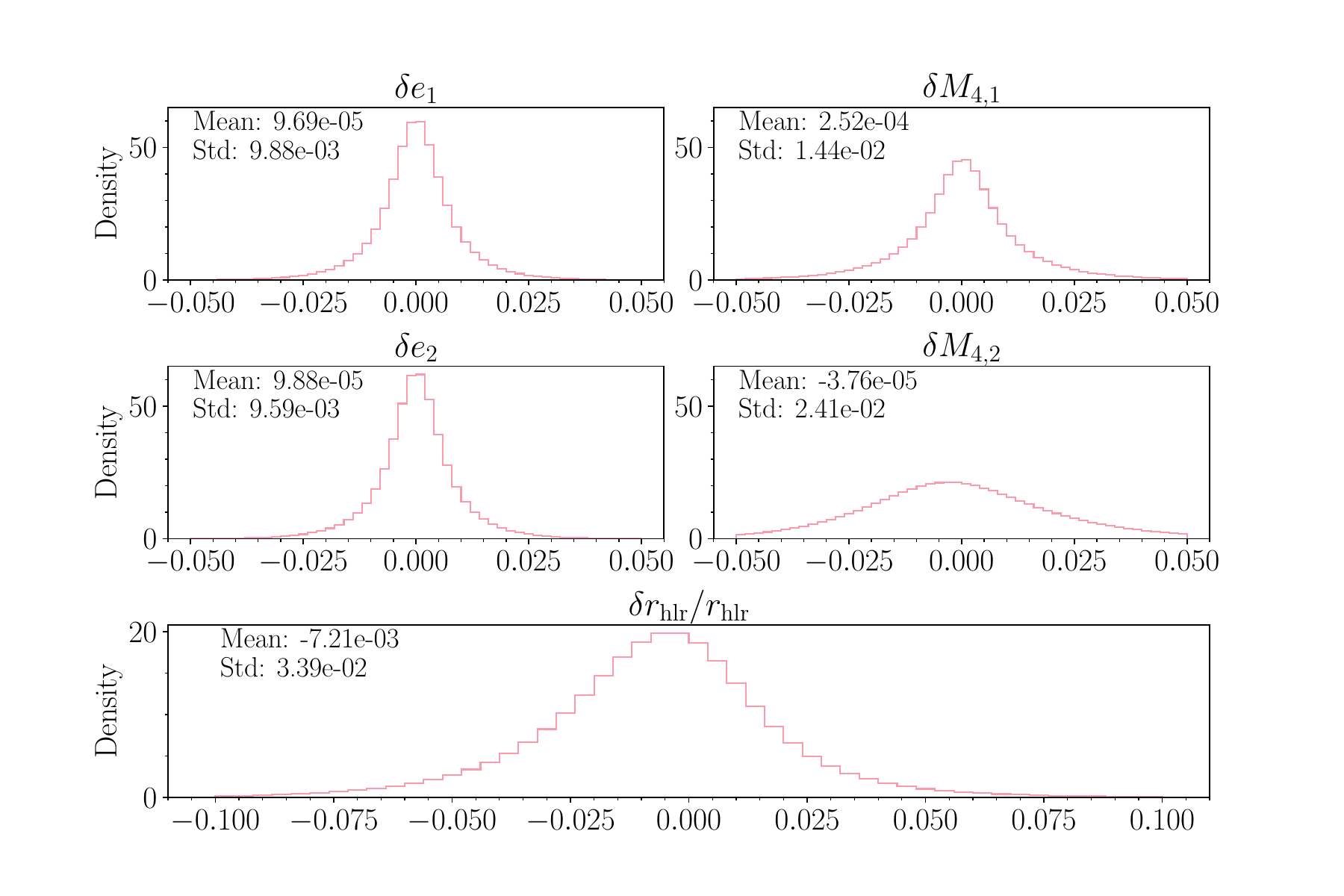}
    \caption{Residuals of the PSF evaluated at validation star positions, shown as histograms. \emph{Top left panels:} First and second ellipticity components. \emph{Top right panels:} First and second fourth-moment components, defined in \cref{eq:fourth_moment_psf}. \emph{Bottom panel:} Normalised size $\delta \tilde{r}_{\rm hlr}^\mathrm{PSF} = (r_{\rm hlr}^* - r_{\rm hlr}^\mathrm{PSF})/r_{\rm hlr}^\mathrm{PSF}$.}
    \label{fig:psf_star_properties}
\end{figure}

To evaluate the performance of the PSF model we estimate observed phenomenological quantities such as size and ellipticities, both on the validation stars and the PSF model interpolated at the position of these stars. Figure~\ref{fig:psf_star_properties} shows the distribution of PSF model residuals at validation star positions. The ellipticity residuals are centred on zero with a small spread. The size residuals, while also peaking at zero, are skewed towards negative values, corresponding to a small overestimation of the PSF size, potentially due to an imperfect star selection. One of the goals of future analyses is to improve upon these residuals. In \cite{jeffersonReanalysisStageIIICosmic2025} the PSF residuals for ellipticities and sizes are shown for other Stage-III surveys, which are similar to ours in width, although the size residuals do not appear skewed. \cite{zhangGeneralFrameworkRemoving2023} introduces higher-order moments to quantify PSF model errors beyond ellipticity and size. We compute these on the PSF and star postage stamps following
\begin{align}
  M_{pq} = \frac{
    \int \mathrm{d} x \, \mathrm{d} y \, u^p v^q \,
        \omega(x, y) I(x, y)
    }{\int \mathrm{d} x \, \mathrm{d} y \,
        \omega(x, y) I(x, y)
    },
    \label{eq:m_pq}
\end{align}
where $(x, y)$ are pixel coordinates, and $(u, v)$ are standardised coordinates, defined in a coordinate system where the second-moment covariance is the identity matrix. This transformation de-correlates higher-order from second-order moments; see \cite{zhangGeneralFrameworkRemoving2023} for details. Additionally, $\omega(x, y)$ is a Gaussian weight function matched to the PSF size, and $I(x,y)$ is the light profile in the image coordinate system. A spin-2 quantity can be built from fourth-order moments as
\begin{align}
\label{eq:fourth_moment_psf}
  M^{(4)} = \left( M_{40} - M_{04} \right) + 2 \textrm{i} \left( M_{13} + M_{31} \right).
\end{align}
As shown in \cref{fig:psf_star_properties}, the distributions of both fourth-order moment components are centred on zero with a small spread, with the second component showing a larger variance than the first one. Overall, the PSF model appears to perform well on validation stars.



One common issue for weak-lensing surveys when choosing calibration stars is the brighter-fatter effect: as more charges are trapped in a given pixel, it gets electrically charged and electrons are repelled to neighbouring pixels, effectively changing the potential lines and making the central pixels smaller. On MegaCam this effect is fairly small \citep{Astier_SNLS_2013,Guyonnet_megacam_2015}, since the thinned CCDs are back-illuminated. The magnitude–radius plot in Fig.~\ref{fig:mag_rad_plot} shows a very constant stellar locus down to $r\sim 16.5$.

\subsection{PSF leakage calibration}\label{seq:object-wise leakage calibration}

In \Cref{sec:shapes} we described the treatment of mean additive and multiplicative biases for shear measurement. Residual PSF leakage can produce additive biases with a particular pattern on the sky corresponding to the underlying PSF field. The amplitude of this leakage can be quantified at the two-point correlation level or on a per-object basis. In \Cref{sec:psf_conta} we will use various validation techniques to assess the amplitude of the PSF leakage at the two-point correlation function level. Here, we review the method to correct PSF leakage at the object level following \cite{liKiDSLegacyCalibrationUnifying2023}. Ultimately, we will apply both the per-object calibration and estimate the residual leakage with two-point statistics. The per-object correction allows for a finer calibration of leakage, accounting for objects' properties in a straightforward manner. The two-point level leakage provides a component averaged estimate of leakage and allows, in addition, to account for PSF residuals.

We define the measured ellipticity of an object with index $i$ as
\begin{align}\label{eq:definition leakage}
    e^\mathrm{obs}_i = \epsilon^\mathrm{s}_i + \gamma_i + \alpha e^\mathrm{PSF}_i,
\end{align}
where $\epsilon^\mathrm{s}_i$ corresponds to the stochastic component of the ellipticity, $\gamma_i$ is the shear component, $e^\mathrm{PSF}_i$ is the interpolated PSF ellipticity at the position of object $i$ and $\alpha$ is the associated leakage coefficient. All three terms are complex spin-2 quantities. We model the leakage $\alpha_i$ as a single scalar for both ellipticity components, which corresponds to the PSF leakage contribution as a simple rescaling of the PSF ellipticity.

\begin{figure}
    \centering
    \includegraphics[width=\linewidth]{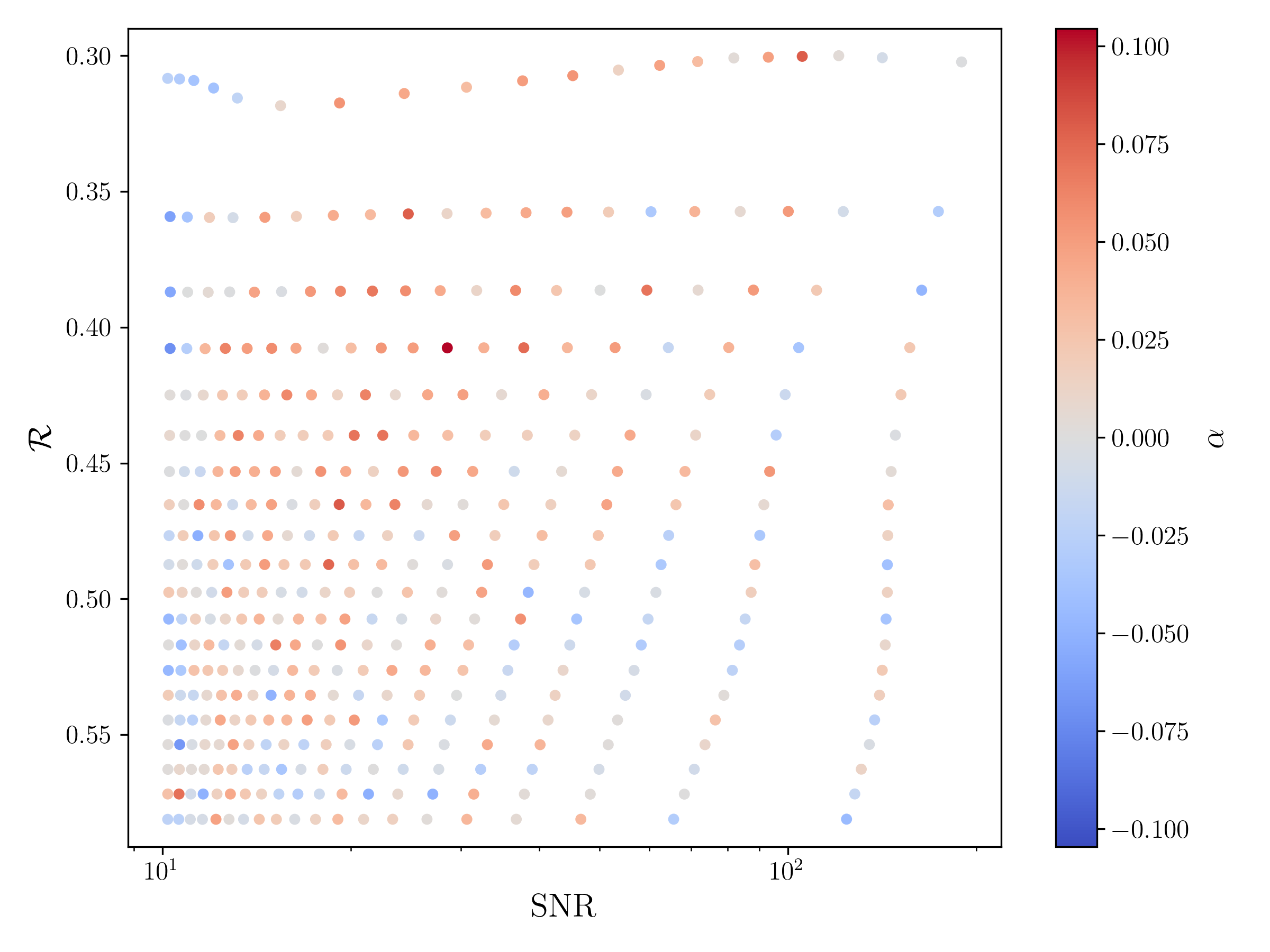}
    \caption{PSF leakage in the measured galaxy ellipticities as a function of SNR and resolution $\size$ \cref{eq:resolution_R}.}
    \label{fig:alpha_leakage_bin}
\end{figure}

The object-wise correction of leakage consists of two consecutive steps to estimate $\alpha_i$ for each object. A first step consists in binning the galaxies by
SNR, $\snr$,
and
resolution, $\size$, with
%
%
\begin{equation}
\size = \frac{r_\mathrm{PSF}}{r_\mathrm{PSF}+r_\mathrm{h}}.
\label{eq:resolution_R}
\end{equation}
We build a $20 \times 20$ grid in ($\snr, \size$) and compute the PSF leakage coefficient $\alpha$ using a linear regression in each bin. Fig.~\ref{fig:alpha_leakage_bin} shows the binned $\alpha$ parameter. Faint and small galaxies tend to have a negative leakage contribution.

Next, we estimate the global leakage trend in the $(\snr, \size)$ plane by fitting the regressed and binned $\alpha$ with the fitting function introduced in \cite{li_kids-legacy_2023}
\begin{align}
    \alpha_\mathrm{trend}(\snr, \size) = a_0 + a_1 \snr^{-2} + a_2 \snr^{-3} + b_1 \size + c_1 \size \snr^{-2}.
\end{align}
We then correct the observed ellipticities by subtracting the fitted leakage,
\begin{align}\label{eq:remove leakage}
    e^\mathrm{obs, tmp}_i = e^\mathrm{obs}_i - \alpha_\mathrm{trend}(\nu_{\mathrm{SNR}, i}, \size_i) e^\mathrm{PSF}_i.
\end{align}
After that first correction, some residual leakage might remain.
To remove this residual leakage further, we perform another suite of linear regressions to estimate, per bin, the residual leakage $\alpha_\mathrm{res}$. We then remove for each object its residual leakage based on the bin it is assigned to, similar to \cref{eq:remove leakage}. The total leakage removed for each object is
\begin{align}
    \alpha_i = \alpha_\mathrm{trend}(\nu_{\mathrm{SNR}, i}, \size_i) + \alpha_{\mathrm{res}, i},
\end{align}
%
where $\alpha_{\mathrm{res}, i}$ is the value of the residual leakage in the $(\size, \snr)$ bin of object $i$. By construction, the measured leakage per bin is close to zero after subtraction. \Cref{sec:psf_conta} will show the impact of this correction on the leakage using different estimators.

\subsection{PSF contamination of galaxy shapes}\label{sec:psf_conta}

As in \cref{seq:object-wise leakage calibration}, we model PSF contamination to galaxy shear estimates as additive, linear biases. In addition to the leakage $\alpha$ term introduced in \cref{eq:definition leakage}, we quantify contributions related to PSF mismodelling with residuals in ellipticity, $\delta \ep = \estar - \ep$, and in normalised size, $\delta \tilde{r}_{\rm hlr}^\mathrm{PSF} = (\Tstar - \Tp)/\Tp$. Here, quantities with the superscript `$*$' denote measurements on stars, while `PSF' refers to PSF model predictions at the position of the star. The star and PSF quantities must be measured with the same methods; otherwise, the residuals mix PSF mismodelling with potential ellipticity and size mismatches between shape measurement techniques. We use adaptive moments with the \texttt{galsim} HSM implementation for both.

Our linear model for the PSF systematic ellipticity is
\begin{align}\label{eq:define_e_sys}
    \esup{PSF, sys} = \alpha \ep + \beta \, \delta \ep + \eta \, \delta \Tp,
\end{align}
The parameters $\alpha$, $\beta$ and $\eta$ modulate the three PSF systematics terms. Following \citep{paulin-henrikssonPointSpreadFunction2008}, the PSF size is redefined as $\delta \Tp = \ep \delta \tilde{r}^\mathrm{PSF}$. The ellipticity factor makes $\delta \Tp$ a spin-2 field, and \cref{eq:define_e_sys} is thus well-defined. Our objective is now to determine $\alpha$, $\beta$ and $\eta$ from two-point correlation functions for the entire galaxy sample.

\subsection{Galaxy-PSF correlation functions}\label{seq:galaxy-PSF correlations}

Galaxy-PSF cross-correlation functions are powerful tools for validating the PSF model and the shape measurement process. These cross-correlations vanish in an ideal case; in practice, they are routinely used to estimate the amplitude of the additive bias in the shear correlation function due to PSF systematics. With the error model from \cref{eq:define_e_sys}, they can be used to distinguish between leakage and mismodelling.

\begin{figure*}[th]
    \centering
    \includegraphics[width=\linewidth]{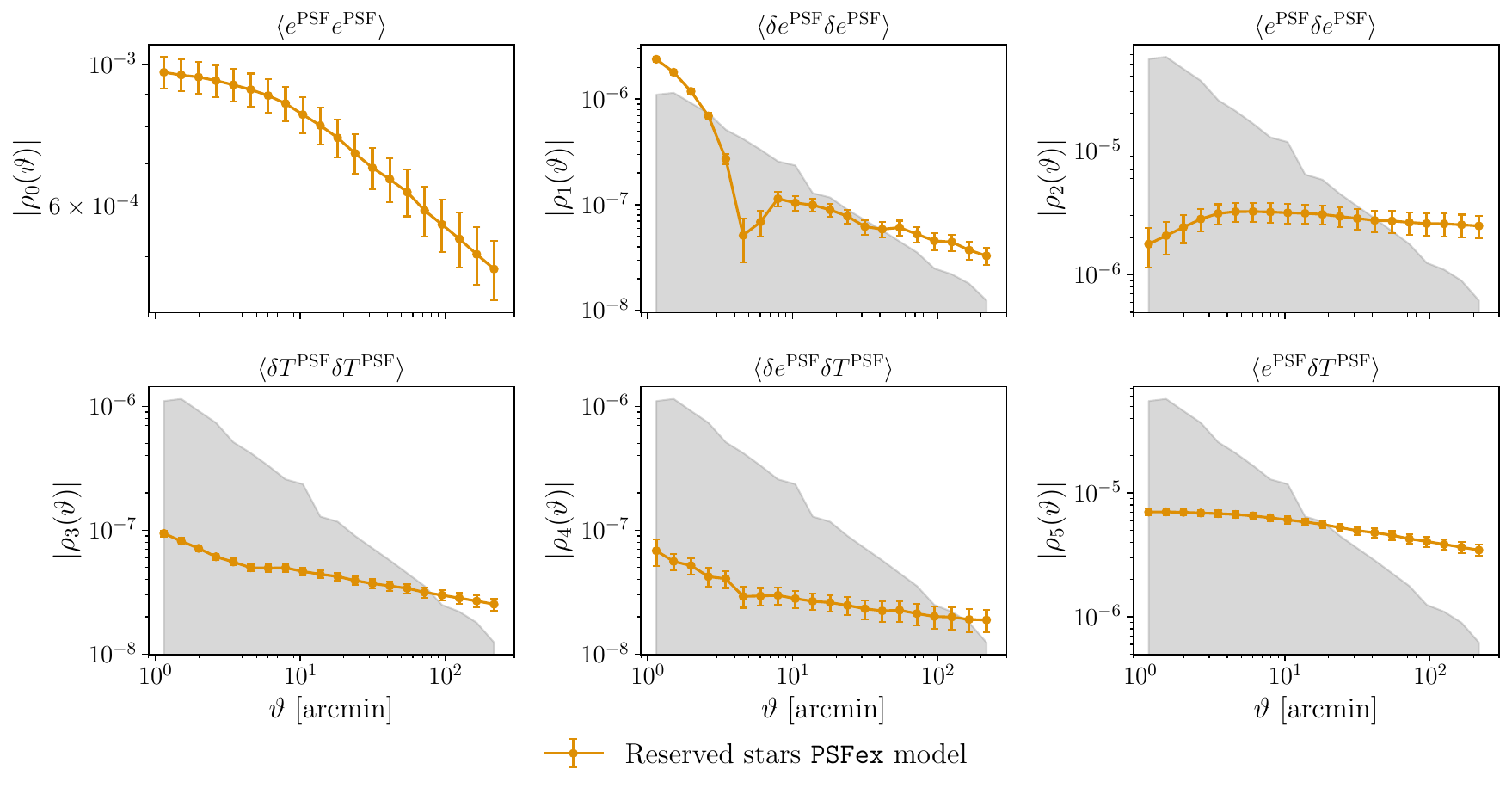}
    \caption{$\rho$-statistics measured on the reserved stars of the fiducial lensing sample. Shaded regions correspond to requirements derived following \cite{mandelbaumFirstyearShearCatalog2018}. Excess on large scales creates PSF leakage quantified in Section~\ref{seq:galaxy-PSF correlations}.}
    \label{fig:rho_stats}
\end{figure*}

A first step is to compute PSF and PSF residual correlations at star positions, known as $\rho-$statistics \citep{roweImprovingPSFModelling2010,jarvisScienceVerificationWeak2016,gattiDarkEnergySurvey2021}. They allow for the comparison of the performance of different PSF models. Here we use them to estimate the amplitude of the leakage bias. Figure~\ref{fig:rho_stats} shows the different $\rho_i$ with $i \in [0,..., 5]$ corresponding to the different cross-correlations in \cref{eq:define_e_sys}. The different $\rho$ correlation functions are:
\begin{align}
\rho_{0}(\theta) &= \left\langle e^{\mathrm{PSF}} e^{\mathrm{PSF}} \right\rangle(\theta); 
& \rho_{1}(\theta) &= \left\langle \delta e^{\mathrm{PSF}} \, \delta e^{\mathrm{PSF}} \right\rangle(\theta); \\[6pt]
\rho_{2}(\theta) &= \left\langle e^{\mathrm{PSF}} \, \delta e^{\mathrm{PSF}} \right\rangle(\theta); 
& \rho_{3}(\theta) &= \left\langle \delta r_{\rm hlr}^{\mathrm{PSF}} \, \delta r_{\rm hlr}^{\mathrm{PSF}} \right\rangle(\theta); \\[6pt]
\rho_{4}(\theta) &= \left\langle \delta e^{\mathrm{PSF}} \, \delta r_{\rm hlr}^{\mathrm{PSF}} \right\rangle(\theta); 
& \rho_{5}(\theta) &= \left\langle e^{\mathrm{PSF}} \, \delta r_{\rm hlr}^{\mathrm{PSF}} \right\rangle(\theta).
\end{align}
The shaded region corresponds to $0.5\delta \xisys{+}$. We estimate the amplitude from $\xi_+(\theta)$ and its SNR following \cite{mandelbaumFirstyearShearCatalog2018}. The $\rho$-statistics associated to leakage tend to exceed those requirements on large scales, and $\rho_2$ shows an excess on small scales. The excess is modelled in the configuration-space cosmological analysis (Paper III). For a comparison to other Stage-III surveys see \cite{jeffersonReanalysisStageIIICosmic2025}.

\begin{figure*}
    \centering
    \includegraphics[width=0.9\linewidth]{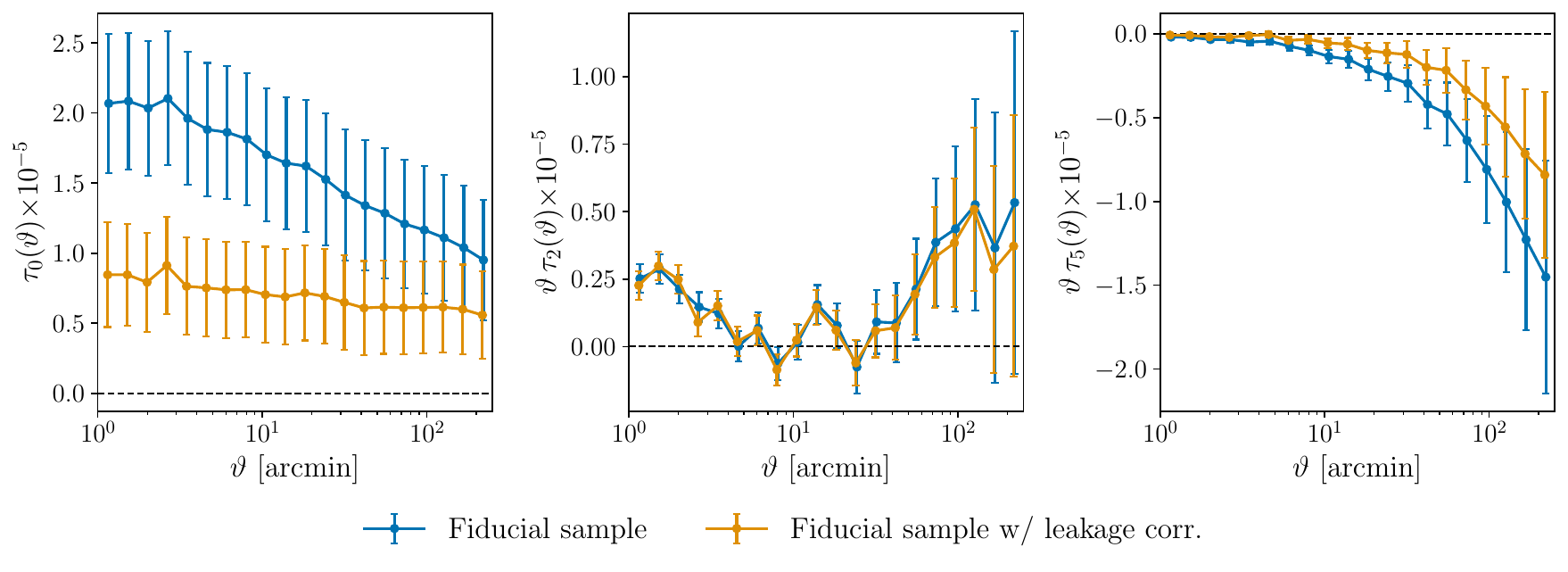}
    \caption{$\tau$-statistics measured on the fiducial weak lensing sample. Blue curve corresponds to the measurement before the object-wise leakage correction is applied. The yellow curve is obtained with the same objects after applying the object-wise correction described in Section~\ref{seq:object-wise leakage calibration}.}
    \label{fig:tau_stats}
\end{figure*}

Despite these excesses, the amplitude of the correlations alone does not reveal the contribution of each $\rho-$statistic to the PSF systematic bias. One must first estimate the values of $\alpha$, $\beta$, and $\eta$.
We perform this estimation by introducing Galaxy-PSF correlations also known as $\tau$-statistics \citep{hamanaCosmologicalConstraintsCosmic2020,giblinKiDS1000CatalogueWeak2021,gattiDarkEnergySurvey2021, zhangGeneralFrameworkRemoving2023,guerriniGalaxyPointSpread2025}. They are defined as
\begin{align}
\tau_{0}(\theta) &= \left\langle e^{\mathrm{gal}} \, e^{\mathrm{PSF}} \right\rangle(\theta); 
& \tau_{2}(\theta) &= \left\langle e^{\mathrm{gal}} \, \delta e^{\mathrm{PSF}} \right\rangle(\theta); \\
 \tau_{5}(\theta) &= \left\langle e^{\mathrm{gal}} \, \delta r_{\rm hlr}^{\mathrm{PSF}} \right\rangle(\theta).
\end{align}
Figure~\ref{fig:tau_stats} shows the three $\tau$-statistics with and without applying the object-wise leakage correction described in Section~\ref{seq:object-wise leakage calibration}. The object-wise correction reduces the amplitude of $\tau_0$ and $\tau_5$, which is expected. $\tau_0$ directly relates to the amount of leakage and $\tau_5$ is strongly correlated with $\tau_0$. This correlation becomes clear in the constraints on $\alpha$, $\beta$, and $\eta$ below.

The measurement of those two types of correlation functions allows us to estimate the parameters of the PSF ellipticity error model $\Omega = (\alpha, \beta, \eta)^T$. Using notations from \cite{guerriniGalaxyPointSpread2025}, we perform it solving:
\begin{align}
    \tau = R \Omega + \Sigma,
\end{align}
where $\tau$ is the flattened data vector of the three $\tau$-statistics, $\Sigma$ denotes the noise term of the estimator and $R$ is a matrix built from $\rho-$statistics.  These quantities are related via the relations
\begin{align}
\tau_{0}(\theta) &= \alpha \, \rho_{0}(\theta) + \beta \, \rho_{2}(\theta) + \eta \, \rho_{5}(\theta); \\[6pt]
\tau_{2}(\theta) &= \alpha \, \rho_{2}(\theta) + \beta \, \rho_{1}(\theta) + \eta \, \rho_{4}(\theta); \\[6pt]
\tau_{5}(\theta) &= \alpha \, \rho_{5}(\theta) + \beta \, \rho_{4}(\theta) + \eta \, \rho_{3}(\theta).
\end{align}

\begin{figure}
    \centering
    \includegraphics[width=\linewidth]{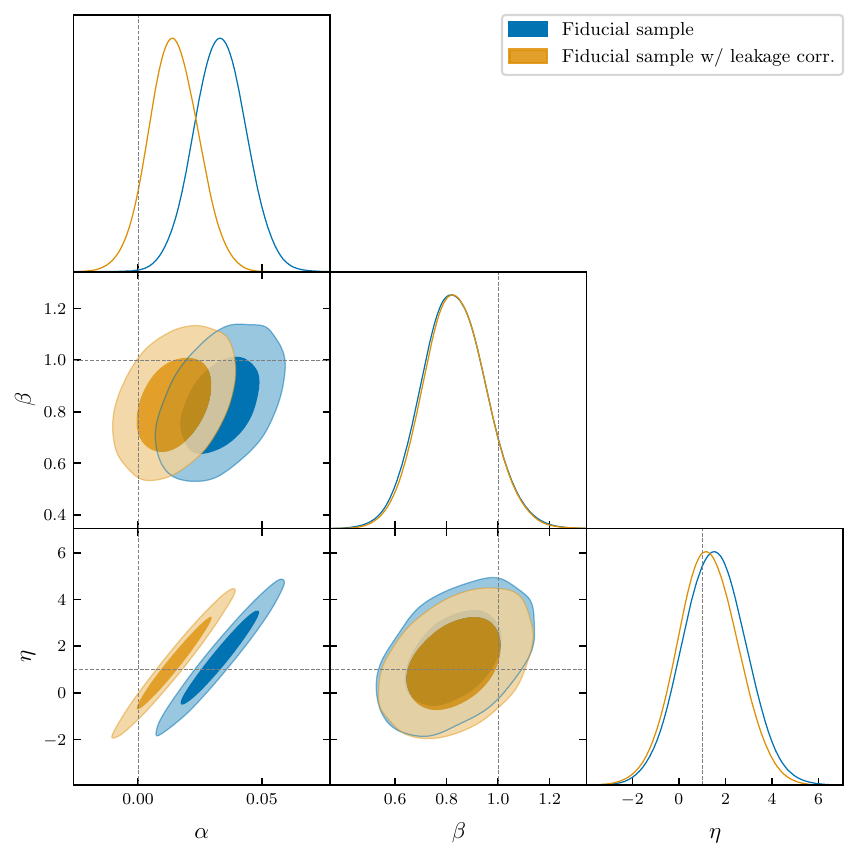}
    \caption{Constraints on PSF parameters $\alpha$, $\beta$ and $\eta$ obtained using a semi-analytical covariance matrix and a least-squares estimator. Dark (light) shaded colours indicate $68.3\%$ ($95.5$\%) confidence regions.}
    \label{fig:psf_leakage_params}
\end{figure}

\begin{figure}
    \centering
    \includegraphics[width=\linewidth]{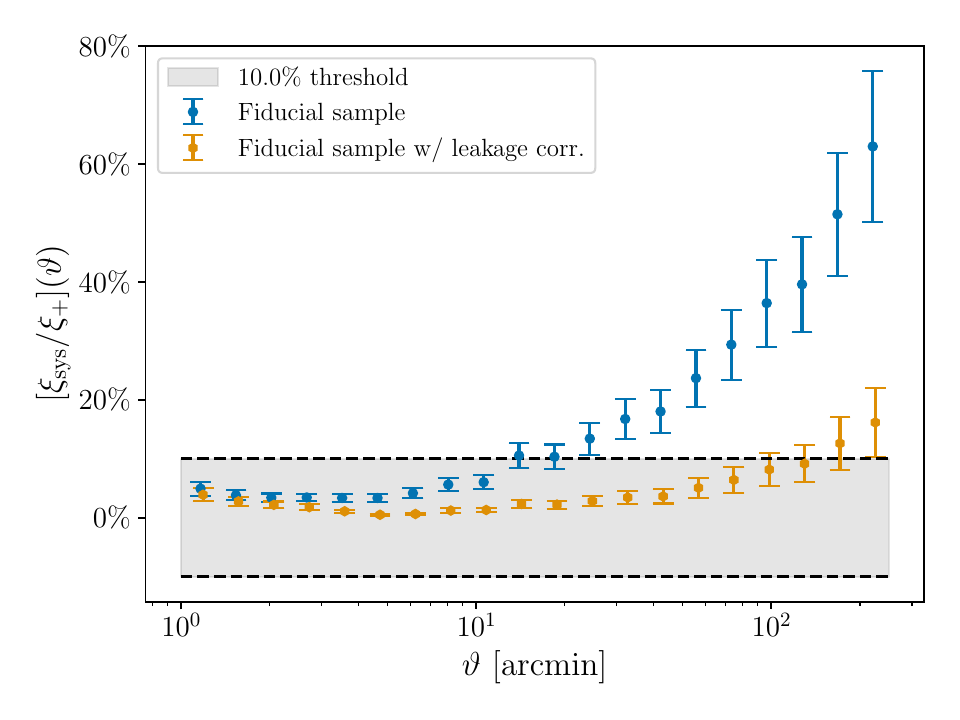}
    \caption{Ratio of the PSF systematic additive bias $\xisys{+}(\theta)$ and $\xi_+(\theta)$ for the object-wise corrected (\emph{blue points}) and uncorrected (\emph{orange}) catalogues.}
    \label{fig:xi_sys_over_xi_plus}
\end{figure}

We apply the methodology developed in \cite{guerriniGalaxyPointSpread2025} and estimate $\Omega$ using a semi-analytical covariance matrix and a least-squares estimator. Constraints on $\Omega$ are shown in Figure~\ref{fig:psf_leakage_params}. The object-wise correction also reduces the $\alpha$ parameter in this case. The $(\alpha, \eta)$-plane shows a strong degeneracy between those parameters. This relates to the multiplication by $\ep$ in the definition of the size residuals. This correlates the leakage and size error terms in such a way that a quantitative measurement of the leakage must be evaluated in the $(\alpha,\eta)$-plane rather than using the 1D posterior on $\alpha$. \\
We propagate these estimates to assess the amplitude of the additive bias due to PSF systematics contributing to the two-point correlation functions $\xi_\pm(\theta)$. We refer to this additive systematic bias as $\xisys{\pm}(\theta)$. It is estimated from the previously established [$\alpha,\beta,\eta$] coefficients as 
\begin{align}
\xi_{\mathrm{PSF, sys}}(\theta) = 
\alpha^{2} \, \rho_{0}(\theta) 
+ \beta^{2} \, \rho_{1}(\theta) 
+ \eta^{2} \, \rho_{3}(\theta) \ +   \\ 2 \alpha \beta \, \rho_{2}(\theta) 
+ 2 \alpha \eta \, \rho_{5}(\theta) 
+ 2 \beta \eta \, \rho_{4}(\theta). \nonumber 
\end{align}
Ideally, this additive contribution should be negligible compared to the cosmological signal. Figure~\ref{fig:xi_sys_over_xi_plus} shows the ratio of the systematic contribution, $\xisys{+}(\theta)$, with respect to the cosmological signal. With a threshold of 10\%, the leakage correction clearly reduces the PSF systematic contribution to $\xi_+(\theta)$ on all scales. Significant leakage contributions still pollute our data beyond $100\arcmin$. We compare the $\xisys{+}$ obtained from the $\rho$ and $\tau$ model with that obtained from $\alpha$ only in Appendix \ref{sec:rho_tau_vs_scale}, finding agreement on large scales. We therefore conclude that the residual PSF contamination is sufficiently mitigated for the cosmological applications targeted in the companion papers by the propagation of the contamination of the PSF systematics signal and scale cuts for large angular separations. 

To assess if other observational properties than the PSF field contaminate the cosmological inference, we empirically test the correlation of ellipticities with various parameters, finding sufficiently low contamination levels in Appendix \ref{append:correl_bias}.

\section{Conclusions}
This paper establishes UNIONS-3500 as a robust UNIONS weak-lensing catalogue designed for cosmology. It is built on the imaging data of \cite{gwynUNIONSUltravioletNearInfrared2025}, follows the methodology of \Axel, and uses the \textsc{ShapePipe} pipeline (\citealp{farrensShapePipeModularWeaklensing2022},\Axel). Its main challenge is the control of shear systematics, especially PSF-related additive contamination, and a central result of this work is to show that these effects are sufficiently characterised and mitigated for the companion cosmic-shear analyses. The catalogue is therefore the foundation for the next generation of UNIONS lensing measurements. This release covers $3{,}648$ deg$^2$ of imaging and yields an effective area $A_\mathrm{eff}=2{,}894~\mathrm{deg}^2$ and effective number density $n_\mathrm{eff}=4.96~\mathrm{arcmin}^{-2}$ for the fiducial v1.4.6.3 catalogue.

In \Cref{sec:data} we described the processing pipeline and its different steps, detailing the data products created in the process, discussing detection, masking strategy and galaxy selection. In \Cref{sec:shapes} we delved into the technical details of the shape measurement and its results, with a presentation of the shear responses and multiplicative biases determined from realistic image simulations. In \Cref{seq:PSF} we lay out the star selection and discuss a dedicated method based on star-star correlations ($\rho$-statistics) and star-galaxy correlation functions ($\tau$-statistics) to evaluate a systematic contamination to the two-point galaxy correlation function $\xisys{\pm}(\theta)$. This allows us to propagate the additional PSF signal into the correlation function and determine scale-cuts to use in a cosmological analysis to keep the PSF contributions below an acceptable threshold. 

To further verify the robustness of the catalogue, we run a suite of null tests presented in \Cref{sec:tan_test} and \Cref{sec:tile_ccd}. The ellipticity distributions in the two core processing units, tiles and CCDs, presented in Figs.~\ref{fig:foc_plane_avg} and \ref{fig:tile_avg} show no correlation with position. Similarly the galaxy-galaxy lensing around tile and CCD centres is consistent with zero.
To verify that bright stars do not leave a coherent imprint on galaxy shapes, we compute the galaxy-galaxy lensing signal around Gaia stars for an independent sample. These two-point correlation functions are consistent with the null hypothesis, indicating that problematic regions have been masked out. In Appendix \ref{append:correl_bias} we show how ellipticities correlate with different galaxy and observational properties, but determine that these produce a negligible contamination at the two-point correlation function level, even for variables which seem to correlate significantly with ellipticity. To connect this work to the different UNIONS lensing papers published so far we detail the versioning process adopted by the collaboration in Appendix \ref{append:cat_versions}, with the key difference being the PSF model used, with \textsc{MCCD} applied to v1.3 and \textsc{PSFEx} used in v1.4. 

In modern weak lensing surveys, the detection of B-modes often serves as a key systematic test to indicate the presence of spurious signals, since no physical effect is known to produce curl-like correlation functions at the precision levels at which Stage-III surveys operate. A dedicated paper (Paper II) explores these B-mode correlations using pure-mode correlation functions $\xi_\pm^{\mathrm{E/B}}(\theta)$, COSEBIS, and harmonic-space angular power spectra $C_\ell^{BB}$. From these correlators, a range in which B-mode contaminations are compatible with 0 is identified, further validating this catalogue for a cosmological analysis.

This paper is the first in a suite of works aimed at using this data to obtain cosmological constraints, in particular on the lensing amplitude $S_8$. Each companion paper contributes to this effort:
\begin{itemize}
\item Paper II presents detailed B-mode calculations, further validating this data-set and exploring the scales on which a robust cosmological analysis is feasible.
\item Paper III shows the first application of this data set to constrain cosmological parameters using cosmic shear measurements.
\item Paper IV presents harmonic-space cosmological constraints from this catalogue.
\item Paper V details the calibration efforts for both shear and redshift, relying on external simulations to validate the methodology and estimate residual biases.
\end{itemize}

While some published papers already made use of this processing, many more scientific questions can be addressed with this data set. 
This catalogue offers strong possibilities of synergies with other surveys in the northern sky. The DESI-DR2 \citep{DESI_DR2_BAO} catalogue will offer unprecedented overlap with UNIONS.
Similarly J-PLUS \citep{JPLUS_Hernandez-Monteagudo_2024} and J-PAS \citep{J_PAS_2014} will measure a large variety of overlapping tracers for which UNIONS can provide detailed lensing masses. 
The \textit{Euclid} DR2 catalogue will provide a large number of objects with shape measurements and slitless spectroscopy on a large overlapping area \citep{euclidcollaborationEuclidPreparationDR12025}, enabling interesting validation efforts.
To extract the full cosmological constraining power of the UNIONS data, photometric redshifts will be added in order to recover the line-of-sight distribution of the sources. This work is ongoing and will result in a full tomographic analysis.

\begin{acknowledgements}
We are honoured and grateful for the opportunity of observing the Universe from Maunakea and Haleakala, which both have cultural, historical and natural significance in Hawai'i. This work is based on data obtained as part of the Canada-France Imaging Survey, a CFHT large program of the National Research Council of Canada and the French Centre National de la Recherche Scientifique. Based on observations obtained with MegaPrime/MegaCam, a joint project of CFHT and CEA Saclay, at the Canada-France-Hawai'i Telescope (CFHT) which is operated by the National Research Council (NRC) of Canada, the Institut National des Science de l’Univers (INSU) of the Centre National de la Recherche Scientifique (CNRS) of France, and the University of Hawai'i. This research used the facilities of the Canadian Astronomy Data Centre operated by the National Research Council of Canada with the support of the Canadian Space Agency. This research is based in part on data collected at Subaru Telescope, which is operated by the National Astronomical Observatory of Japan.
Pan-STARRS is a project of the Institute for Astronomy of the University of Hawai'i, and is supported by the NASA SSO Near Earth Observation Program under grants 80NSSC18K0971, NNX14AM74G, NNX12AR65G, NNX13AQ47G, NNX08AR22G, 80NSSC21K1572 and by the State of Hawai'i.
This work was made possible by utilising the CANDIDE cluster at the Institut d’Astrophysique de Paris. The cluster was funded through grants from the PNCG, CNES, DIM-ACAV, the Euclid Consortium, and the Danish National Research Foundation Cosmic Dawn Center (DNRF140). It is maintained by Stephane Rouberol. We gratefully acknowledge support from the CNRS/IN2P3 Computing Center (Lyon - France) for providing computing and data-processing resources needed for this work.
The authors acknowledge the use of the Canadian Advanced Network for Astronomy Research (CANFAR) Science Platform operated by the Canadian Astronomy Data Centre (CADC) and the Digital Research Alliance of Canada (DRAC), with support from the National Research Council of Canada (NRC), the Canadian Space Agency (CSA), CANARIE, and the Canada Foundation for Innovation (CFI).
LWKG thanks the University of Edinburgh School of Physics and Astronomy for a postdoctoral Fellowship.
CD and MK acknowledge support from the Agence Nationale de la Recherche (ANR-22CE31-0014-01) TOSCA project.
FHP acknowledges support from CNES.
MJH acknowledges support from NSERC through a Discovery Grant.
LVW acknowledges support from NSERC through a Discovery Grant. 
We would like to thank our external blinding coordinator, Koen Kuijken.
\end{acknowledgements}

\bibliographystyle{aa}
\bibliography{unions-2d-cosmic-shear-bass,unions-2d-cosmic-shear-temp} 

\begin{appendix}

\section{Star and PSF-related maps}

\begin{figure*}[t]
    \centering
    \includegraphics[trim={3cm 0 3cm 0},clip,width=0.48\linewidth]{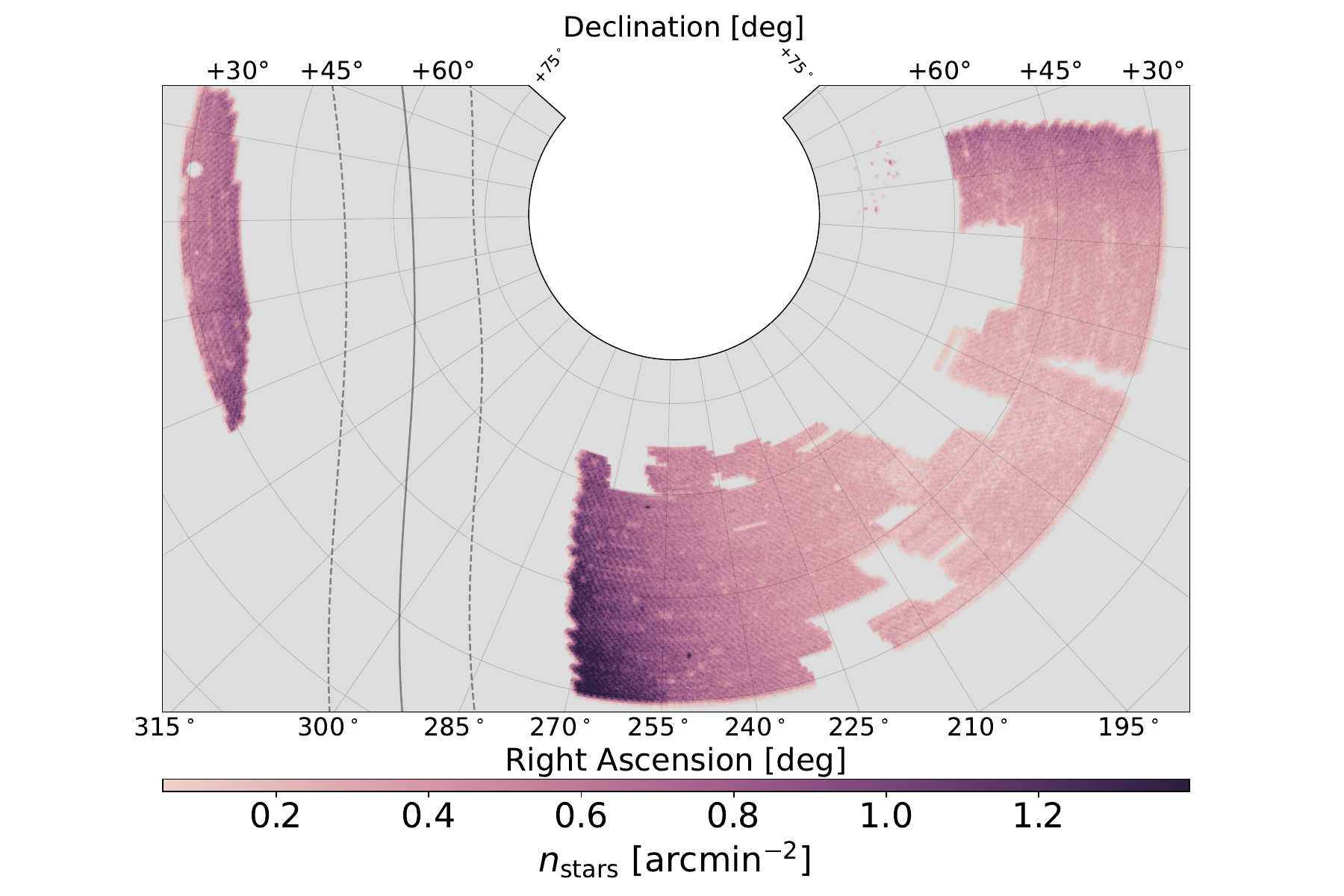}
    \includegraphics[trim={3cm 0 3cm 0},clip,width=0.48\linewidth]{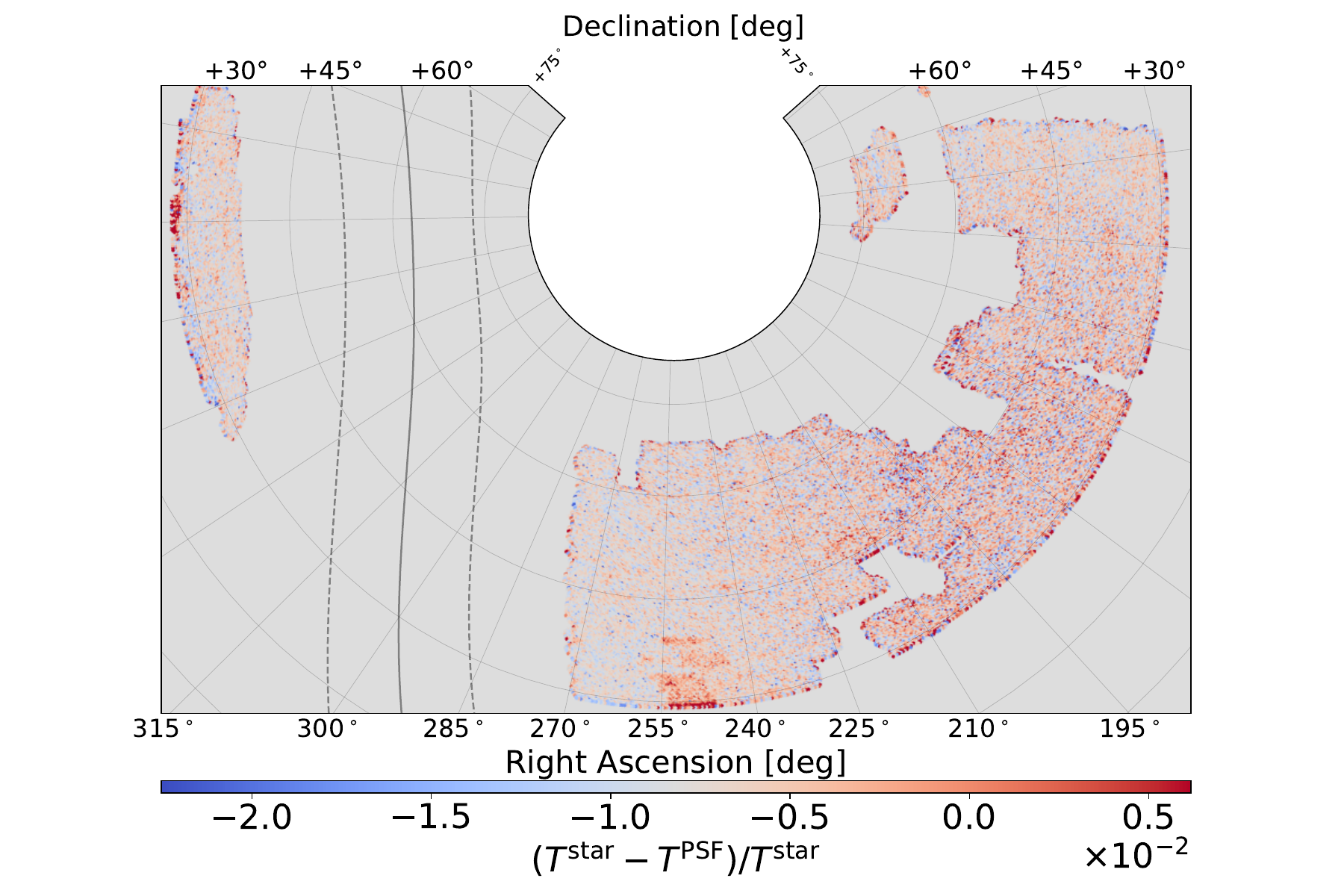}
    \caption{\emph{Left panel:} Distribution of stars detected in the survey. The density increases towards the galactic plane as expected. This map appears correlated with the PSF residual map on the right. \emph{Right panel:} Distribution of star size residuals from the PSF model estimated on reserve stars. Both plots are pixelized with a \texttt{HEALPix} size of \texttt{NSIDE}=512}.
    \label{fig:star_density_residuals}
\end{figure*}

\begin{figure*}[t]
    \centering
    \includegraphics[trim={3cm 0 3cm 0},clip,width=0.48\linewidth]{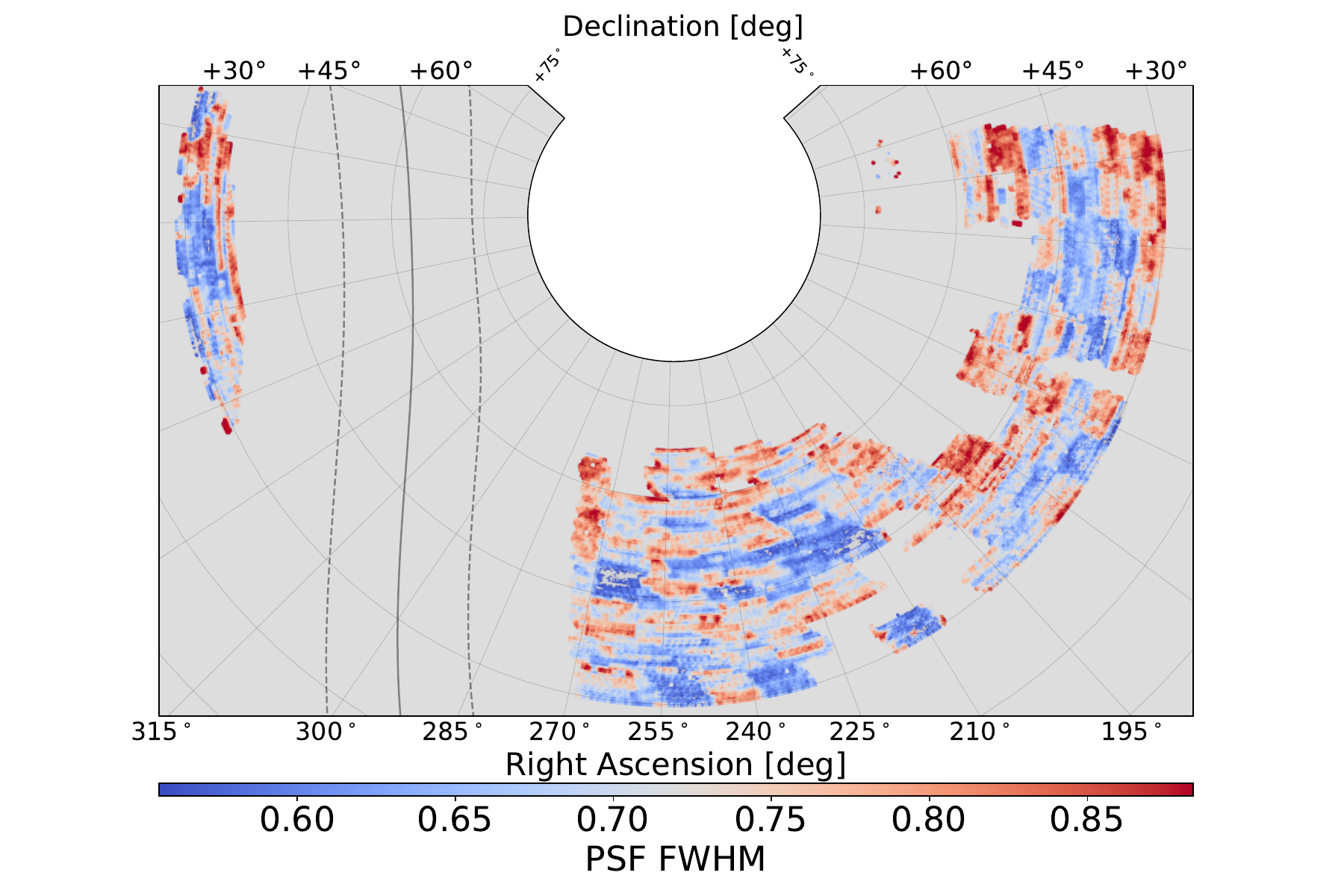}
        \includegraphics[trim={3cm 0 3cm 0},clip,width=0.48\linewidth]{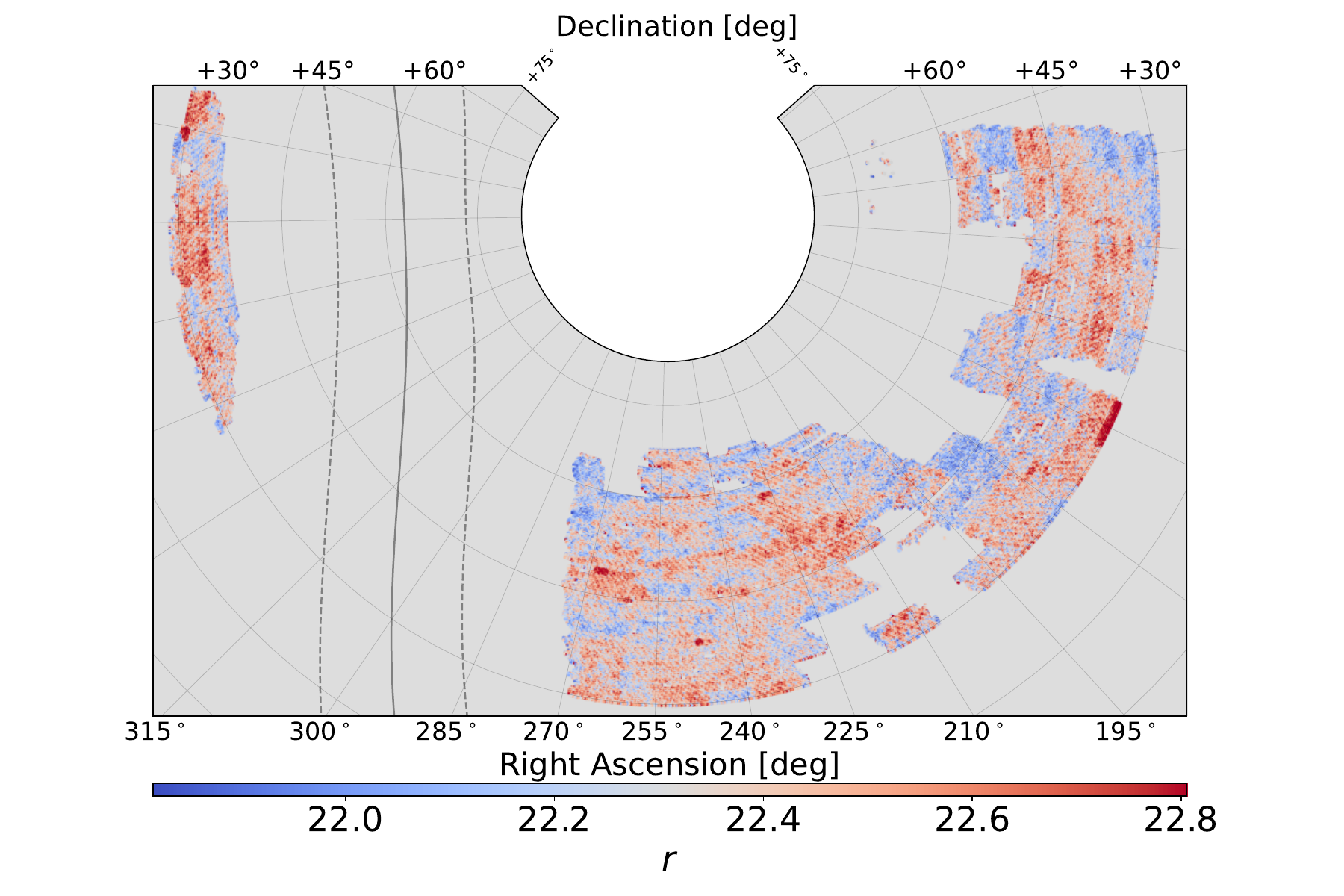}

    \caption{\emph{Left panel:} Spatial map of the PSF size evaluated at the galaxy positions. \emph{Right panel:} The distribution of mean magnitudes in pixels with \texttt{NSIDE}=512. Some structure can be seen which appears to correlate with the PSF FWHM map.}
    \label{fig:mag_fwhm_psf_map}
\end{figure*}

\begin{figure*}[t]
    \centering
    \includegraphics[trim={3cm 0 3cm 0},clip,width=0.48\linewidth]{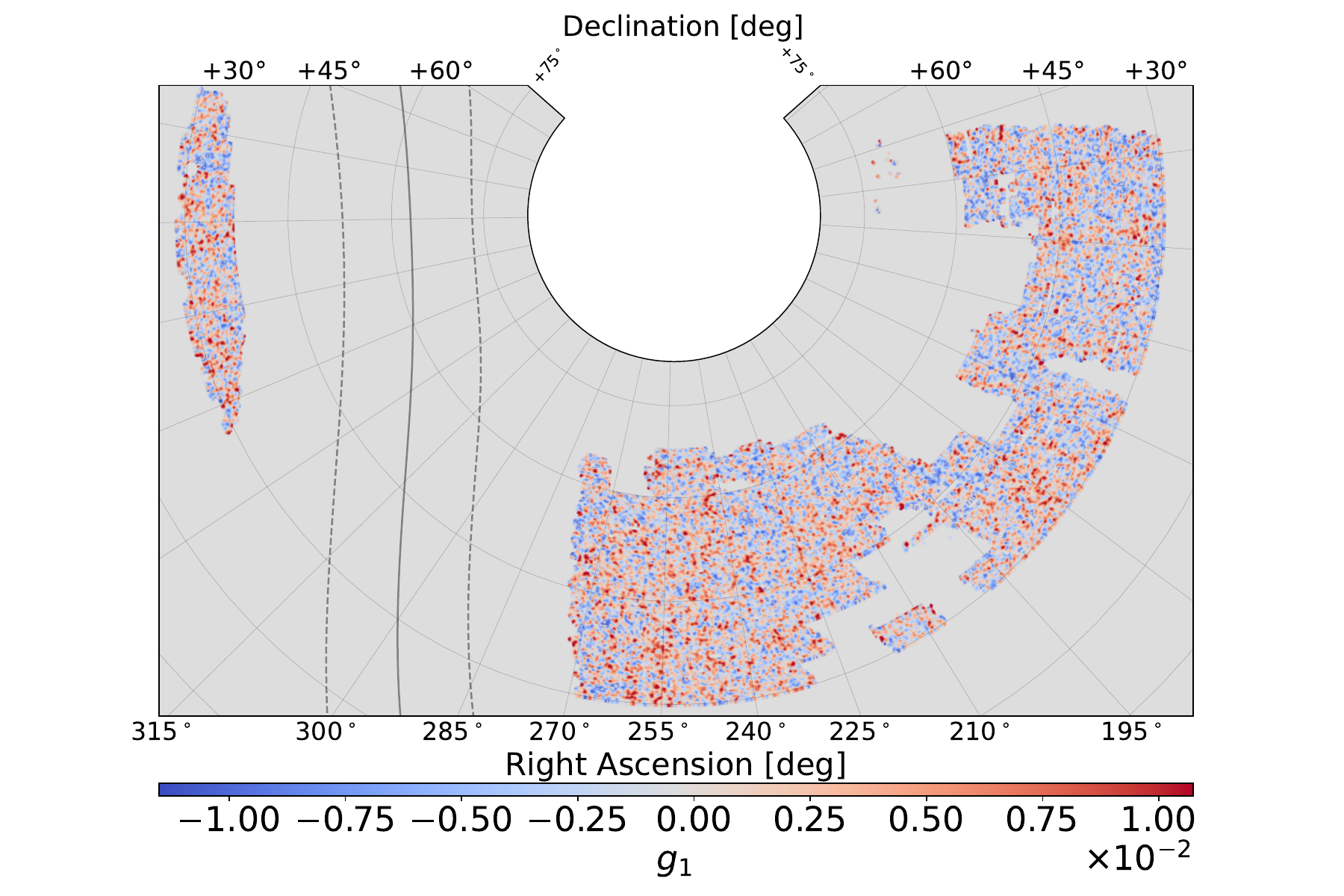}
    \includegraphics[trim={3cm 0 3cm 0},clip,width=0.48\linewidth]{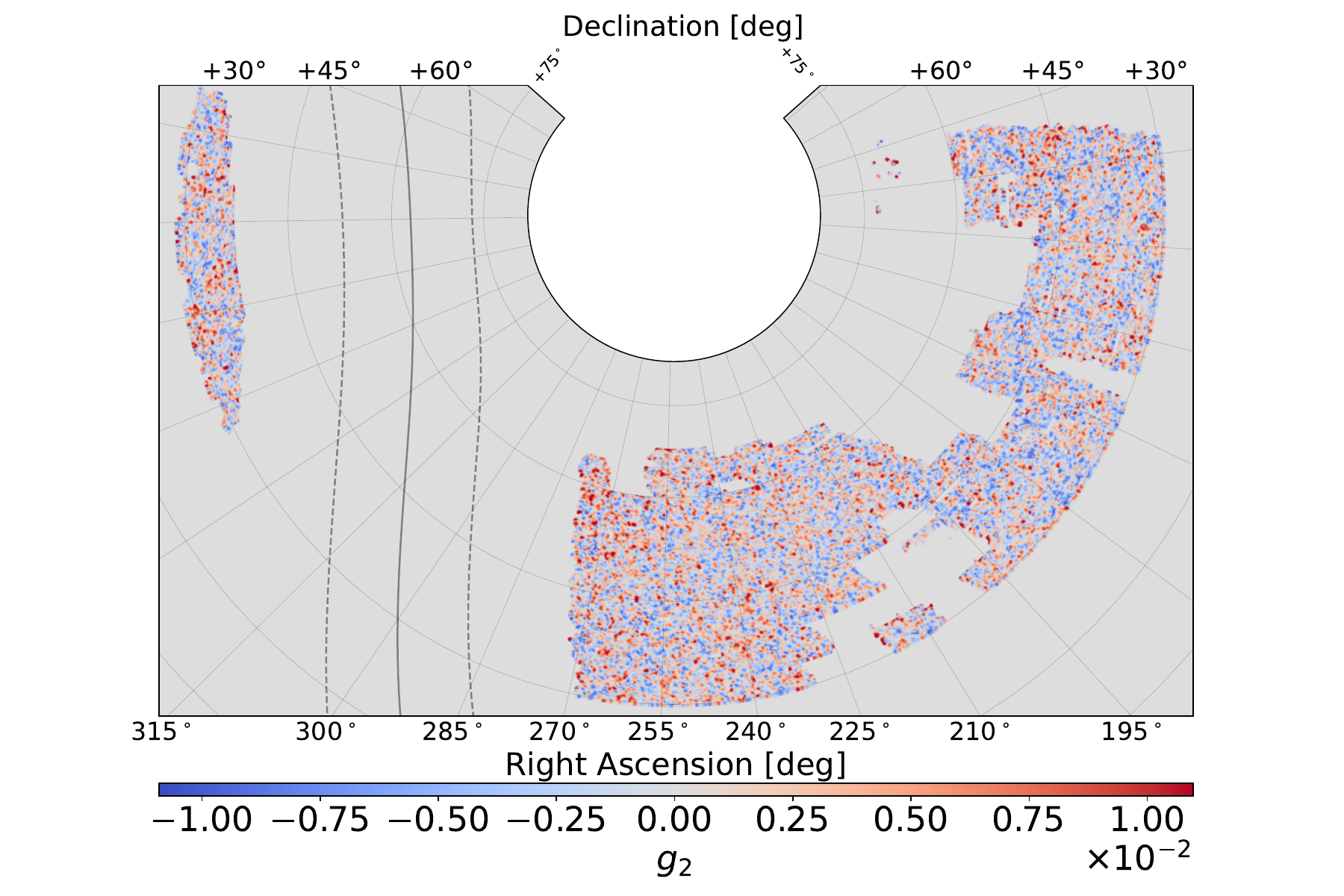}
    \caption{Average $g_1$ and $g_2$ reduced shear components as measured by \texttt{ngmix} binned in pixels of size \texttt{NSIDE}=512. No spatial correlation appears in this plot indicating an absence of obvious spatially distributed additive systematics.}
    \label{fig:e1_map}
\end{figure*}

To assess the homogeneity of the data quality and to verify the absence of any obvious spatially correlated systematics, we present the sky-coverage of different quantities from our final selected galaxy and star samples. Note that a Gaussian smoothing is applied which causes some masked area in the data to be filed in. Fig.~\ref{fig:star_density_residuals} is showing the stellar densities; a clear pattern emerges with denser star densities located towards the galactic plane and a sparser coverage in the regions further out. Interestingly, as shown in the same figure, this central region is also where the PSF model is less precise. This indicates that our PSF model is limited by the available number of stars in the field.

The distribution of the PSF FWHM, see Fig.~\ref{fig:mag_fwhm_psf_map}, shows a lot of structure. Although the PSF size is averaged over different exposures taken over a number of nights, its spatial correlation can be interpreted as the varying atmosphere even though the observation strategy was performed with an emphasis on data homogeneity. The PSF size map appears to anti-correlate with the average magnitude map of Fig.~\ref{fig:mag_fwhm_psf_map}: In regions with better seeing, a larger number of small (and faint) galaxies are selected due to the cut in galaxy size with respect to PSF size.

As is discussed in Appendix \ref{append:correl_bias}, no trend of the average ellipticity is expected with respect to any observational quantity in an unbiased measurement. It is therefore a common diagnostic to plot the mean ellipticity distributions on a map to capture spatially systematic deviations from a pure noise realisation, verifying that no spurious correlation exists with a potentially untested phenomenological variable. The distributions of the $g_1$ and $g_2$ ellipticity components as measured by \texttt{ngmix} are shown in Fig.~\ref{fig:e1_map}. No spatial pattern seems to emerge. The spread is consistent across the two components indicating an absence of any obvious spatially correlated additive systematic effect.

\section{Masking partition} 
\begin{figure*}
    \centering
    \includegraphics[width=0.84\linewidth]{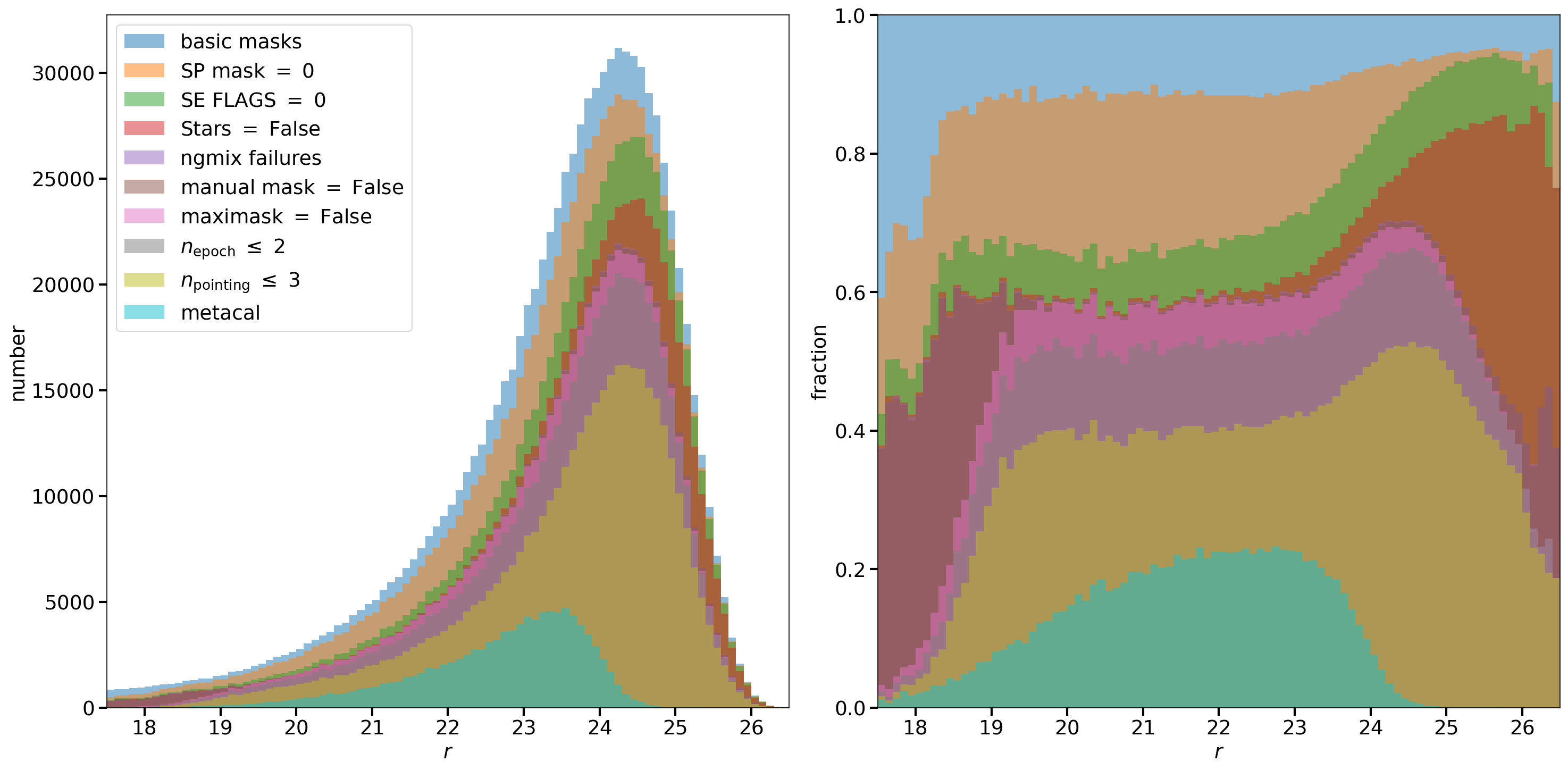}
    \caption{Histogram of the $r$-band magnitude of detected objects with various cumulative selection criteria: %
     all unmasked objects with valid shapes, chip defects and cosmic rays removed (blue); %
     with the bright foreground objects removed (orange);
     with the \texttt{SExtractor} flag=0 applied (green);
     with stellar haloes from \texttt{THELI} removed (red);
     with \texttt{ngmix} failures removed (purple);
     with additional foreground objects removed (brown);
     with \texttt{maximask} applied (pink);
     with $n_\textrm{exp} \geq 2$ (grey); %
     with $n_\textrm{pointing} \geq 3$ (green); %
     with size and SNR cuts performed during metacalibration (cyan).}
     \label{fig:cuts_histo}
\end{figure*}

As laid out in \cref{sec:masking}, we employ a very conservative detection threshold, and then apply successive selection cuts. The various successive masking steps are  depicted in Fig.~\ref{fig:cuts_histo} as a function of magnitude. The different cuts are described in \cref{sec:masking}. Different masks have different magnitude dependencies. We briefly reiterate what each mask means and describe their magnitude dependence.

The \textit{basic masks} sample corresponds to the first MegaCam processed masks. The IMA\_FLAGS=0 cut removes objects at the very bright end of the magnitude distribution, which is expected as these take out highly luminous features such as foreground galaxies and diffraction spikes. 
The \texttt{SExtractor} FLAGS = 0 choice, which was presented in \cref{sec:m_bias} to lead to a multiplicative bias, excludes objects in the brighter range $r\in[18,24]$. Fainter objects are not large enough to trigger the deblending. Next, the stars mask affects objects at all magnitudes as any kind of object can fall into the stellar haloes. Further, the failures of \texttt{ngmix} are mainly at the fainter end. This can be caused by a bad deconvolution or a non-converging fit. The manual mask is an additional mask which affects things like large foreground objects, this is implemented in the THELI pipeline. It is redundant with the previous $\textsc{ShapePipe}$ mask barely affecting any objects. The \texttt{MaxiMask} removes very bright objects ($r<20$), as it mainly focuses on capturing cosmic rays and satellite trails. The epoch and pointing masks are approximately independent of magnitude. Finally the \texttt{Metacalibration} mask removes all objects with $r>24.5$, mainly through the SNR cut. This leaves a distribution peaking around $r\approx$ 23.5.

\section{Comparison of $\rho$ and $\tau$ derived leakage vs. scale-dependent $\alpha$ }\label{sec:rho_tau_vs_scale}

\begin{figure}
    \centering
    \includegraphics[width=1\linewidth]{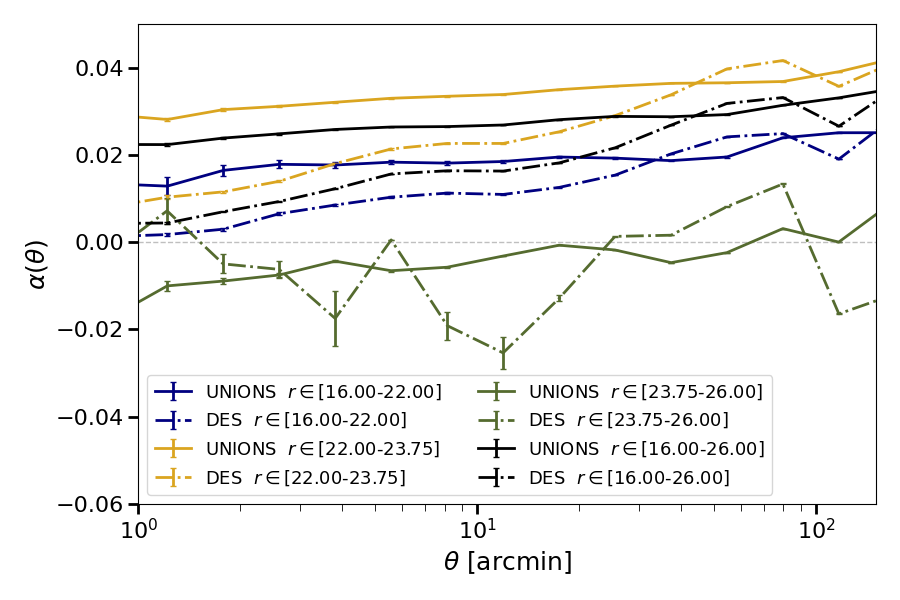}
    \caption{Scale-dependent PSF-leakage parameter $\alpha(\theta)$ for UNIONS and DES-Y3 in magnitude bins.}
    \label{fig:scale_dependent_leakage}
\end{figure}

\begin{figure}
    \centering
    \includegraphics[width=\linewidth]{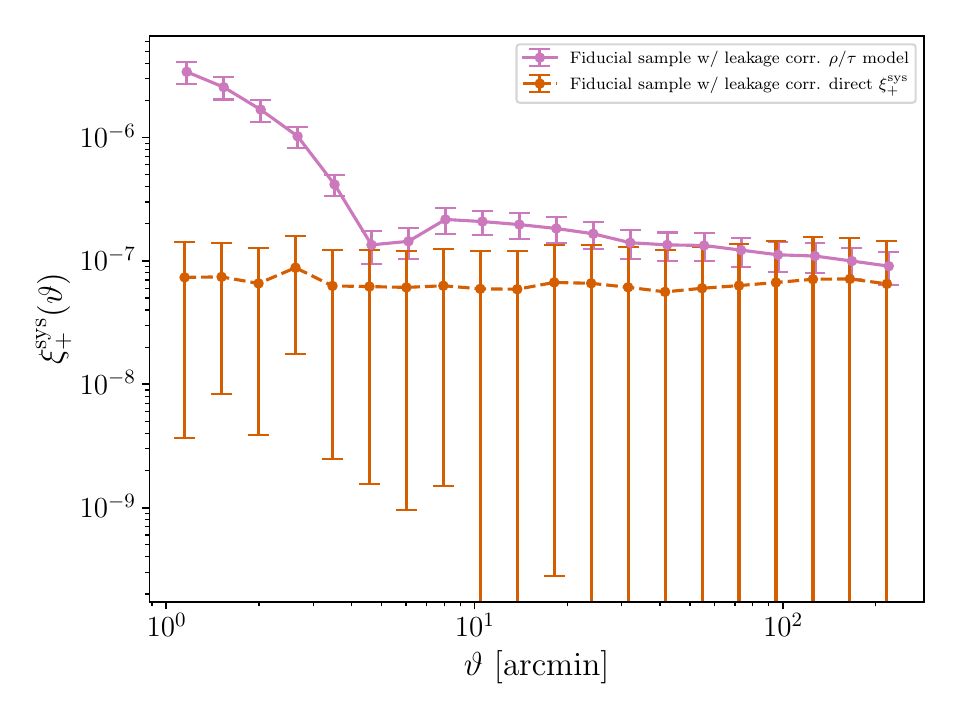}
    \caption{Comparison of the PSF systematic additive bias estimated from the $\rho$ and $\tau$ statistics and from the scale-dependent leakage only.}
    \label{fig:xi_sys_comparison}
\end{figure}

While the $\rho$ and $\tau$-statistics framework for estimating PSF contribution has recently become the standard methodology, we want to compare it to a previously used method referred to as scale-dependent leakage. 
Omitting the $\beta$ and $\eta$ PSF residual terms from \cref{eq:define_e_sys}, the scale-dependent leakage is estimated from $\tau_0(\theta)$ and $\rho(\theta)$ as
\begin{align}
    \alpha(\theta) = \frac{\tau_0(\theta)}{\rho_0(\theta)}.
\end{align}
This function was introduced in \cite{baconJointCosmicShear2003} as a PSF diagnostic function.
Following Section~\ref{seq:galaxy-PSF correlations}, the PSF systematic additive contribution to the two-point correlation function is
\begin{align}
    \xisys{\pm}(\theta) = \alpha^2(\theta)\rho_{0}(\theta) = \frac{\tau_0^2(\theta)}{\rho_0(\theta)}.
\end{align}
Figures~\ref{fig:scale_dependent_leakage} and \ref{fig:xi_sys_comparison} show, respectively, the scale-dependent leakage and the resulting systematic contamination. 
The scale-dependence of the leakage parameter is very mild, remaining mostly constant across scales. This estimator also shows a reduced systematic contribution from leakage after the object-wise correction. Figure~\ref{fig:xi_sys_comparison} compares the predicted systematic additive bias estimated with the scale-dependent leakage and from $\rho-$ and $\tau-$statistics of Section~\ref{seq:galaxy-PSF correlations}. The two estimates match on large scales. The discrepancy on small scales is due to the unmodelled $\beta$ term in the scale-dependent estimator. The scale-dependent estimator is therefore incomplete, but yields a consistent leakage amplitude and impact on the two-point correlation function on very large scales.

In Fig.~\ref{fig:scale_dependent_leakage}, we show the scale-dependent leakage for UNIONS and DES-Y3 for the total samples and with the galaxies split into three magnitude bins. The galaxy ellipticities have not been corrected for leakage but have the bin-related shear response applied. Two important features immediately appear, which are a different scaling with $\theta$ and varying amplitudes for the different magnitude bins. On small scales, the total DES-Y3 leakage is close to 0, which is consistent with the global small $\alpha$ value estimated from the $\rho$ and $\tau$ statistics in \cite{gattiDarkEnergySurvey2021}. The UNIONS scale-dependent leakage is almost constant on all scales, and the DES-Y3 leakage matches the UNIONS one above 60 arcmin. 
When comparing to previous versions of the \textsc{ShapePipe} catalogue, we realised that the object selection can have a strong effect on the global leakage. This is evidenced in Fig.~\ref{fig:alpha_leakage_bin}, where $\alpha$ correlates with SNR and size-ratio. The origin of this dependency is unclear but it is important to consider the object property distributions when comparing global leakage values. It is particularly puzzling that the leakage is not monotonous in brightness.

\begin{figure}
    \centering
    \includegraphics[width=1.1\linewidth]{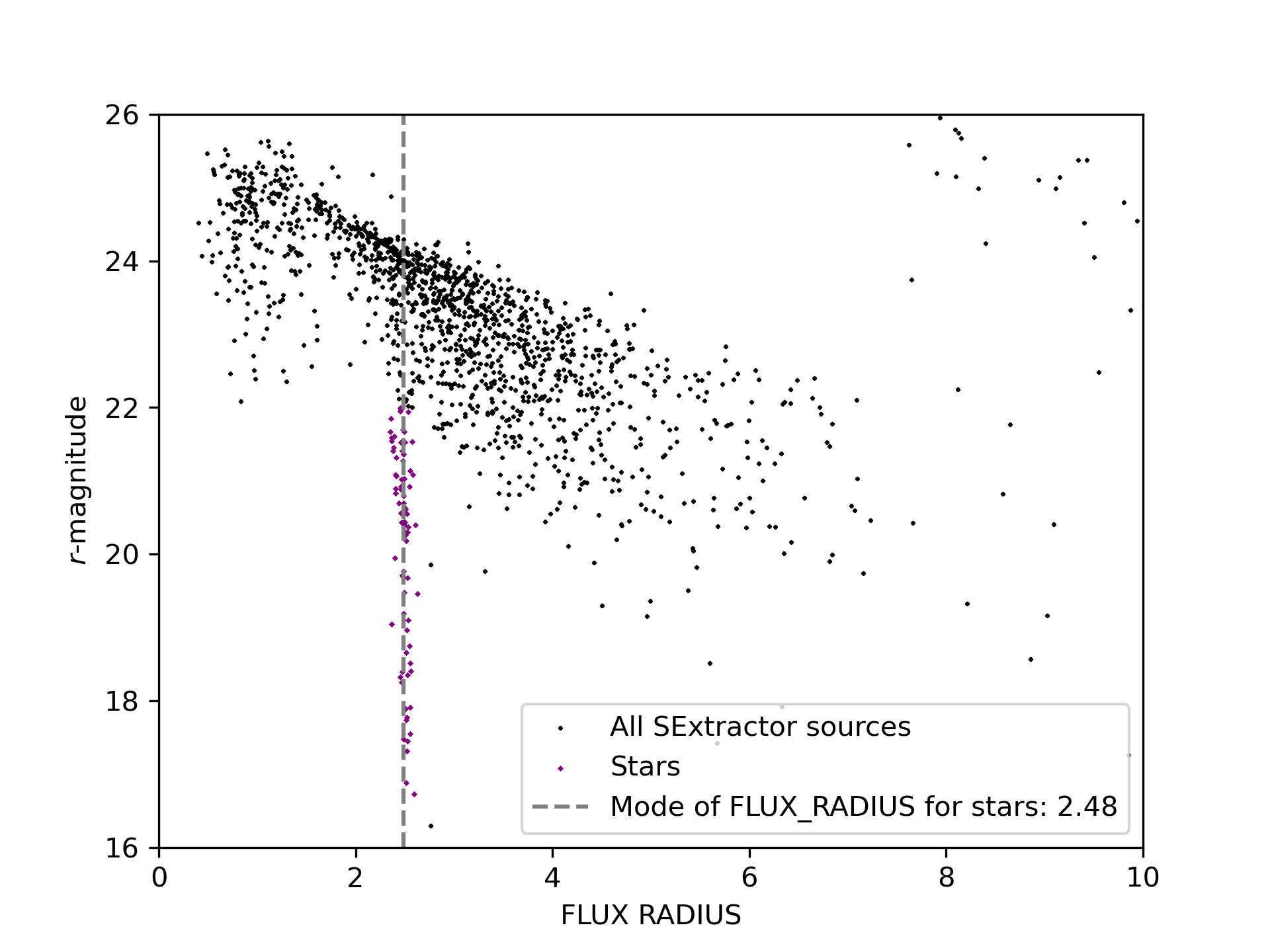}
    \caption{Size-magnitude distribution of \texttt{SExtractor} detected sources on a tile. The violet points are the selected stars following the stellar locus criteria.}
    \label{fig:mag_rad_plot}
\end{figure}

\section{Ellipticity-density PSF contamination}

\begin{figure}
\includegraphics[width=\linewidth]{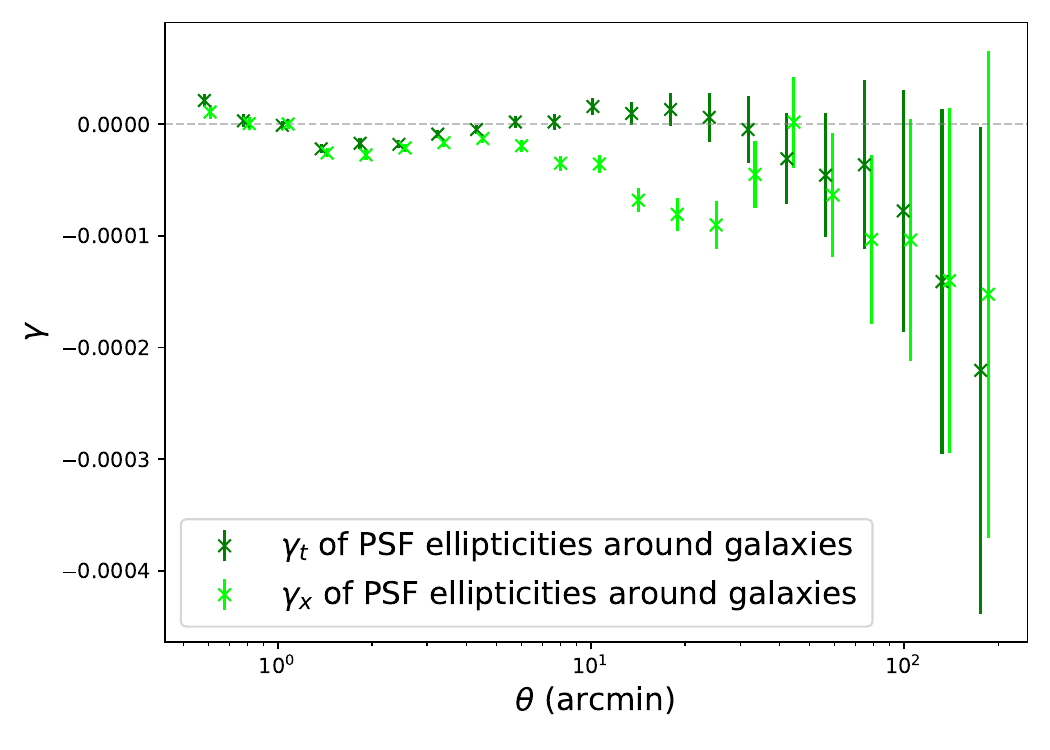}
\caption{Tangential and cross-component shear of UNIONS PSF stars around galaxies.}
\label{fig:gamma_gal_stars_PSFval}
\end{figure}

In \cite{zhangPointSpreadFunction2024}, a framework was developed to capture the additive bias caused by PSF-leakage for position-shape correlation functions based on the model from \cite{paulin-henrikssonPointSpreadFunction2008}. Formally, the 
$\lambda$-statistics are defined as 
\begin{align}
    \lambda_1
        & = \left\langle
                \varepsilon\tang^\PSF \, n
            \right\rangle ;
    \nonumber \\
    \lambda_2
    & = \left\langle
            \frac{\delta T^\PSF} {T^\PSF} \varepsilon\tang^\PSF \, n
        \right\rangle;
    \nonumber \\
    \lambda_3
    & = \left\langle
            \delta \varepsilon\tang^\PSF \, n
        \right\rangle .
    \label{eq:lambda_123}
\end{align}
The density tracer $n$ represents any set of positions, in this case we use galaxies.  These correlation functions cannot be understood as a null test since their results will most likely deviate from 0 due to correlations in the PSF field. In \cite{zhangPointSpreadFunction2024} it is shown how these functions allow us to quantify additive PSF biases in density-ellipticity tracers. This framework was applied to this catalogue in \cite{Hervas_Peters_IA_2024} to quantify the additive biases to the integrated projected position-shape correlation function $w_{g+}$. In Fig.~\ref{fig:gamma_gal_stars_PSFval}, the galaxy position-star ellipticity correlation functions are shown for the tangential and cross components. These correlation functions are very complex to interpret and their deviation from 0 would not be a problem if the leakage parameter $\alpha$=0. For the contamination to a typical galaxy-galaxy lensing type measurement in UNIONS see \cite{zhangGeneralFrameworkRemoving2023}.

\section{Tangential shear tests} \label{sec:tan_test}

\begin{figure*}
    \centering
    \includegraphics[width=0.48\linewidth{}]{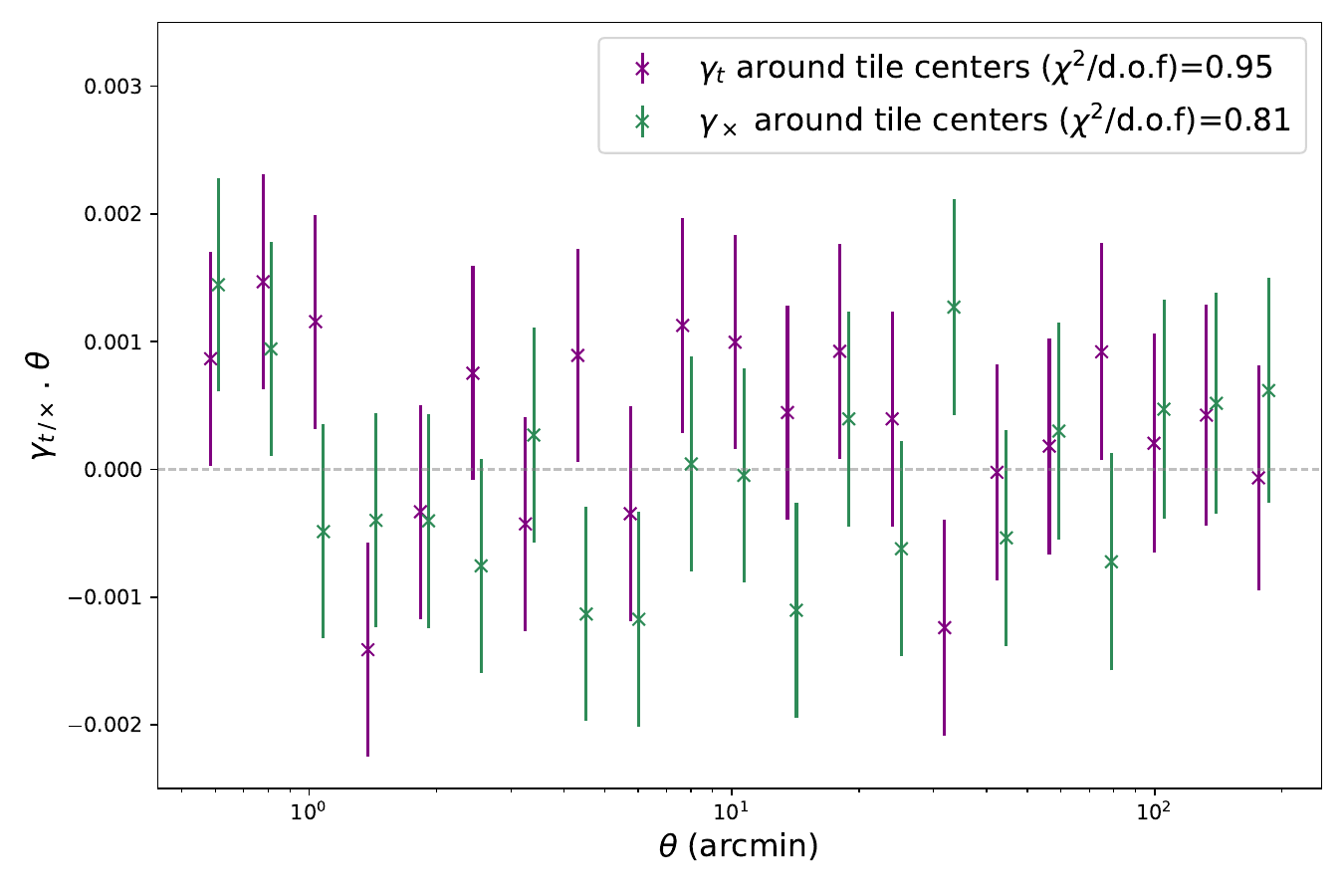}
    \includegraphics[width=0.48\linewidth{}]{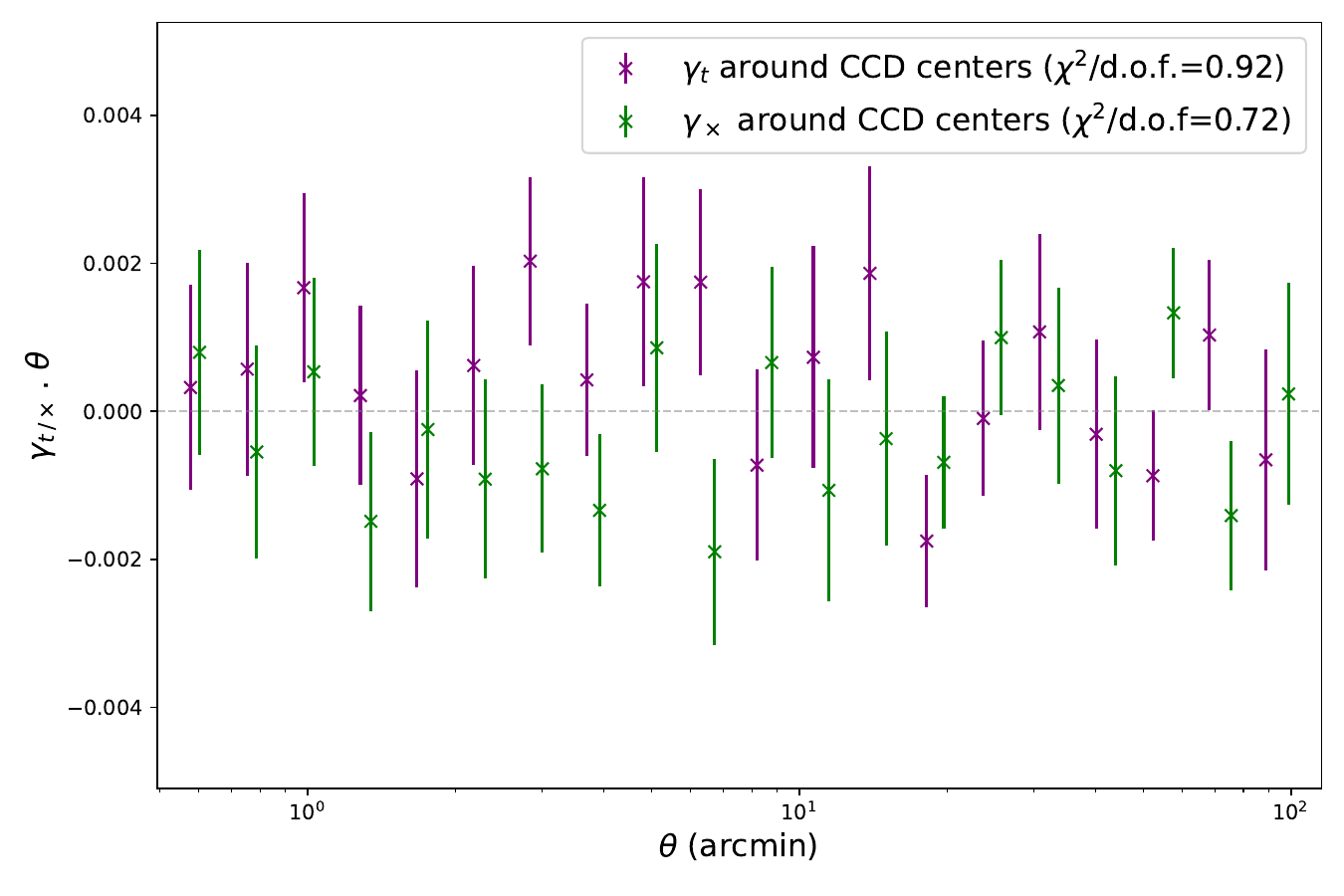}
    \caption{Galaxy ellipticity around tile centres (\emph{left panel}) and CCD centres (\emph{right}). }
    \label{fig:ggl_tile_ccd}
\end{figure*}

Two important shear-position correlation-functions tests are routinely performed to verify the absence of strong systematic radial trends in the data: galaxy ellipticities around processing-unit centres and stars.
Figure~\ref{fig:ggl_tile_ccd} shows $\gamma_\textrm{t}$ and $\gamma_\times$ around tile and CCD centres. The reduced $\chi^2$ values are consistent with unity, and no spatial pattern is visible, indicating an absence of radial additive systematics at the tile and CCD level. The error bars only have the shot-noise contribution for the tile-centred correlation functions; the CCD-centred ones use jackknife to account for survey-wide variability.

\begin{figure}
\includegraphics[width=\linewidth]{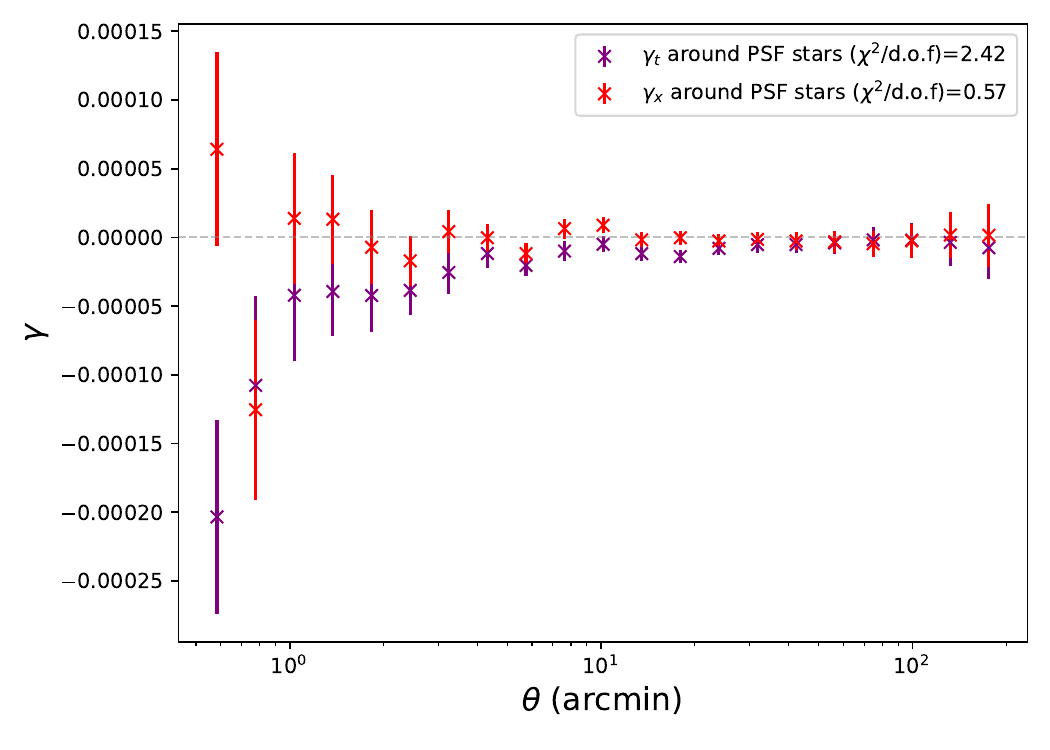}
\caption{Tangential and cross-component shear of galaxies around UNIONS stars. The slight preference for a radial pattern (i.e. negative $\gamma_t$) is not tied to a particular magnitude range.  }
\label{fig:gamma_gal_stars_PSFval_}
\end{figure}

\begin{figure}
\includegraphics[width=\linewidth]{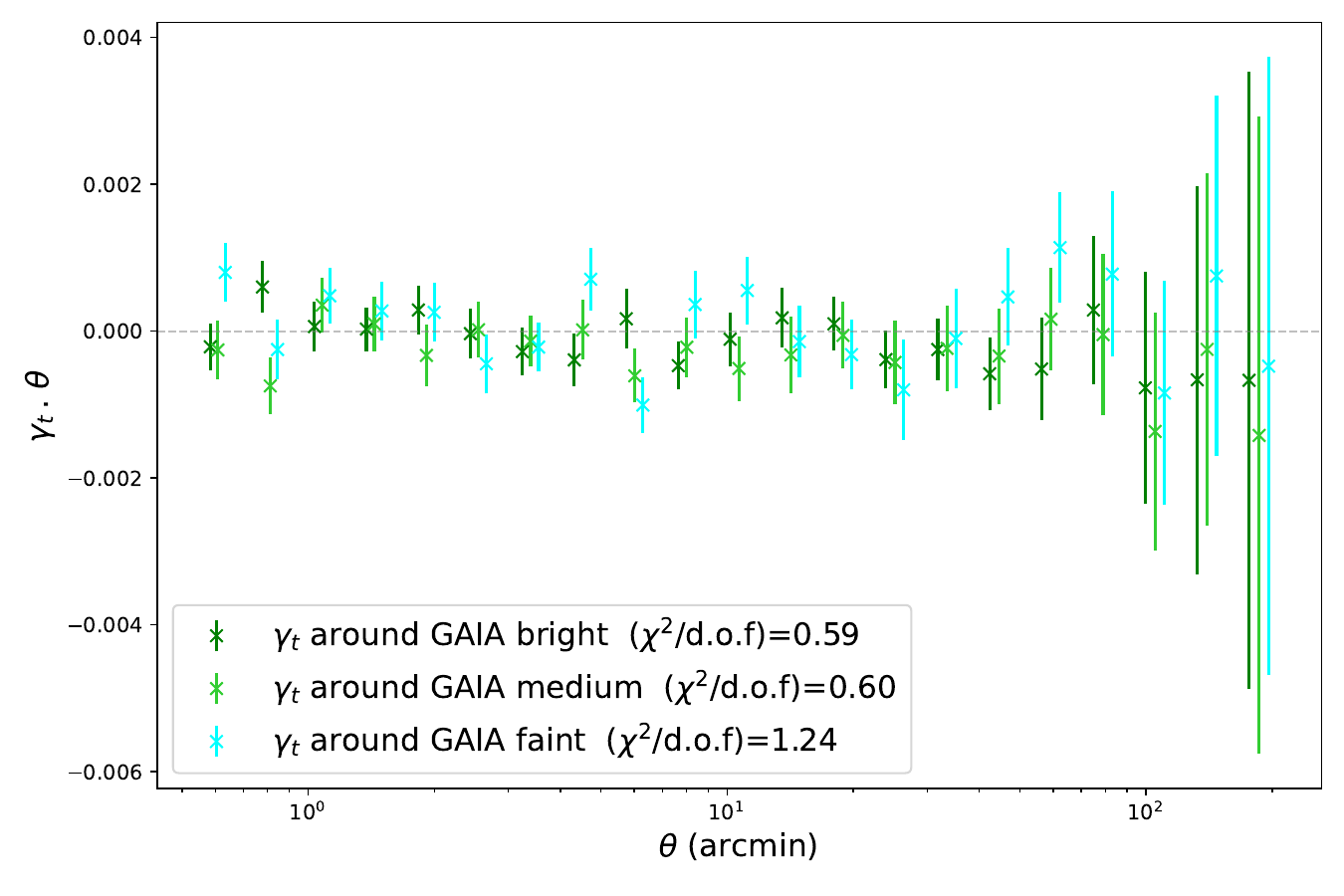}

\includegraphics[width=\linewidth]{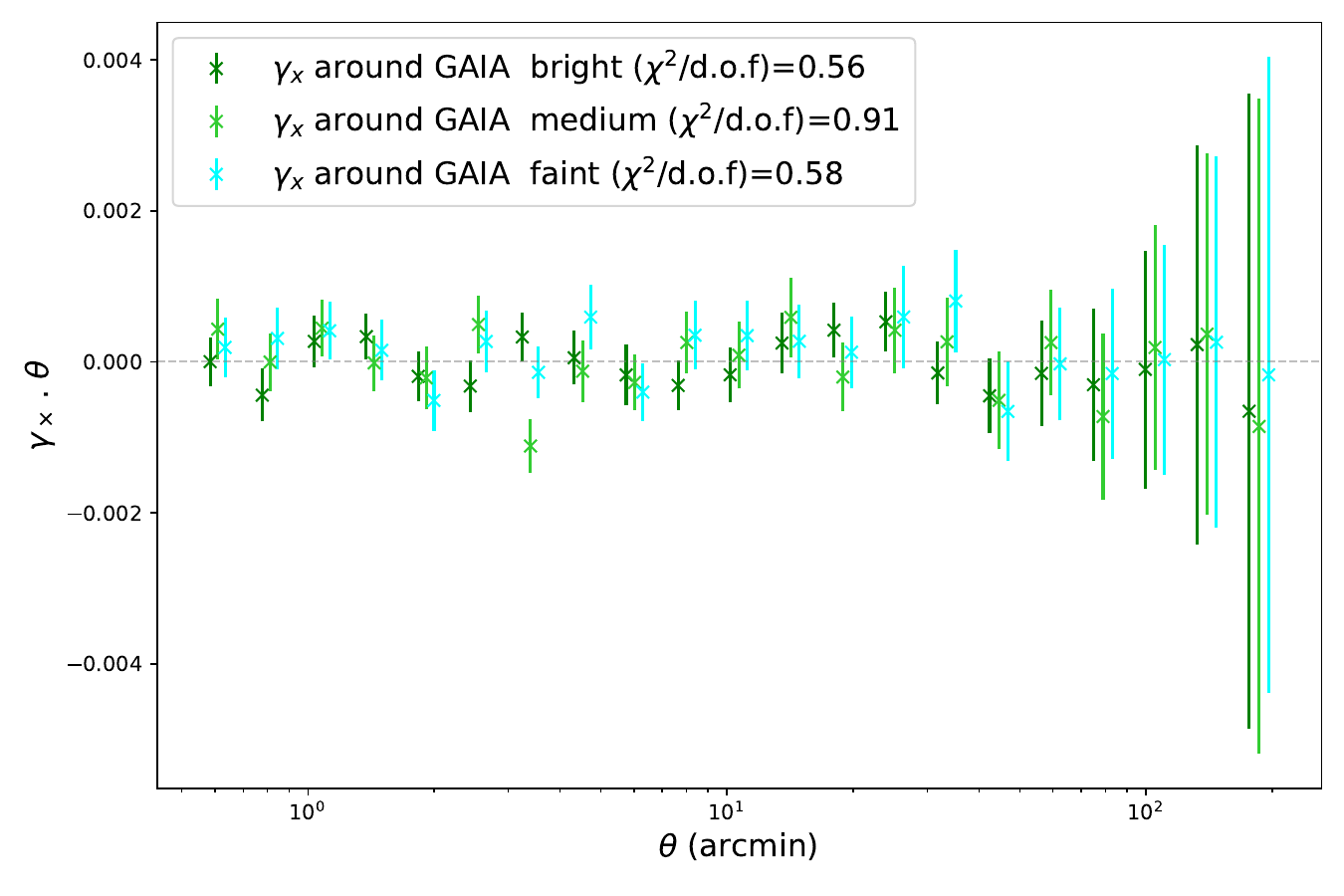}

\caption{Tangential (\emph{top panel}) and cross-component (\emph{bottom}) shear of galaxies around Gaia stars.}
\label{fig:gammat_gal_stars_GAIA}
\end{figure}

A further validation test is whether any unsubtracted light from bright stars coherently affected galaxies during the shape measurement process. In Figs.~\ref{fig:gamma_gal_stars_PSFval_} and \ref{fig:gammat_gal_stars_GAIA} we compute the tangential and cross signals around Gaia \citep{GAIA_DR3_2021} selected stars and around our own PSF selected stars. 

To test the different stellar regimes, we split the Gaia sample into the three $g$-band magnitude bins [3,17.9,19.2,20]. All reduced $\chi^2$ are consistent with 1 indicating no clear systematic offset from 0. The correlation functions and the associated reduced $\chi^2$ are shown in Fig.~\ref{fig:gammat_gal_stars_GAIA}. Visual inspection also suggests that no particular trend appears on a specific scale or between the different magnitude bins.  

We repeat this exercise around UNIONS PSF validation stars. The star selection is described in \cref{sec:star_sel} and illustrated in Fig.~\ref{fig:mag_rad_plot}. The shear around this sample is shown in
\Cref{fig:gamma_gal_stars_PSFval_}. The $\gamma_\times$ test is compatible with the null hypothesis. The tangential shear has a preference for negative values (i.e., radial alignments) up to approximately 30 arcmin, with a reduced $\chi^2$ of 2.53 when considering all scales. We tested binning the stars in magnitude, and the slight preference for a radial alignment was present across all magnitudes. Since this does not appear around Gaia stars, although both samples cover similar magnitude ranges, the first hypothesis coming to mind is that some galaxies made it into the star sample causing a spurious signal. The presence of spurious galaxies in the star sample would cause a tangential alignment resulting in a positive $\gamma_\textrm{t}$, making it a poor hypothesis to explain the observed negative trend. We found no change in significance when varying the masking around stars. This small excess therefore remains an open question.  

\section{Tile and CCD diagnostics} \label{sec:tile_ccd}

\begin{figure*}
    \centering
    \includegraphics[width=0.9\linewidth]{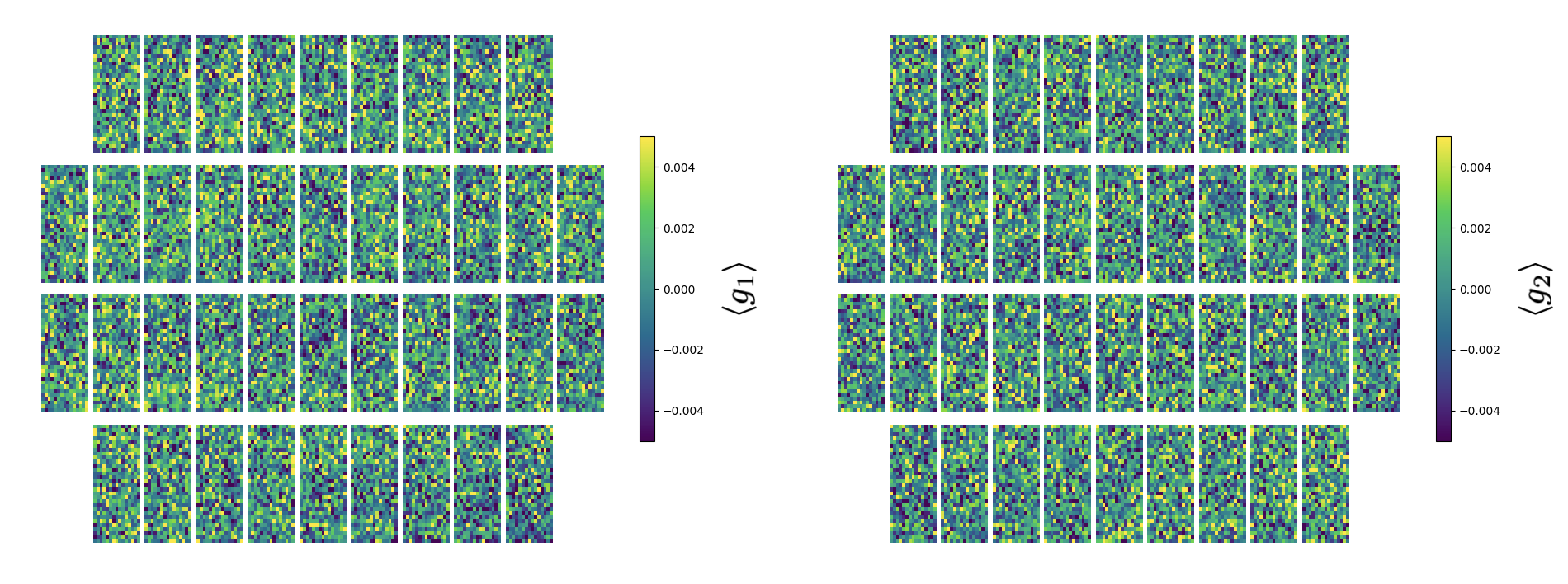}
    \caption{Average ellipticity components (\emph{left panel:} $g_1$; \emph{right:} $g_2$), as a function of position on the focal plane. Due to the complex dither pattern, each galaxy appears at multiple positions on the focal plane. The spread is small and no pattern emerges.  }
    \label{fig:foc_plane_avg}
\end{figure*}

\begin{figure*}
    \centering
    \includegraphics[width=0.45\linewidth{}]{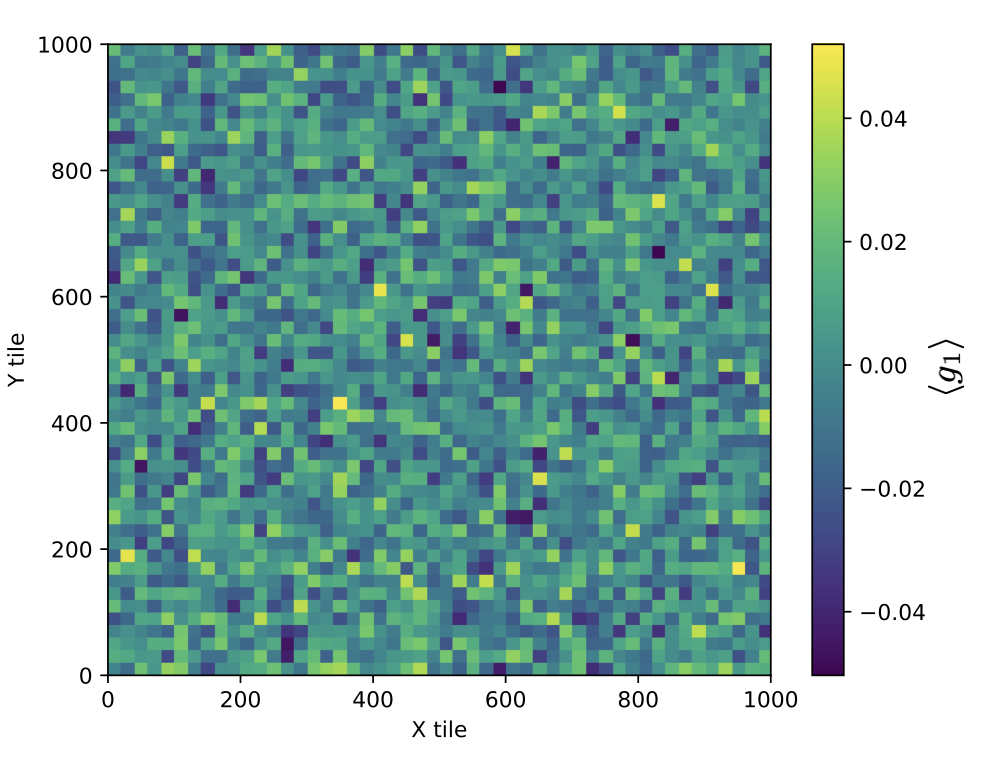}
    \includegraphics[width=0.45\linewidth{}]{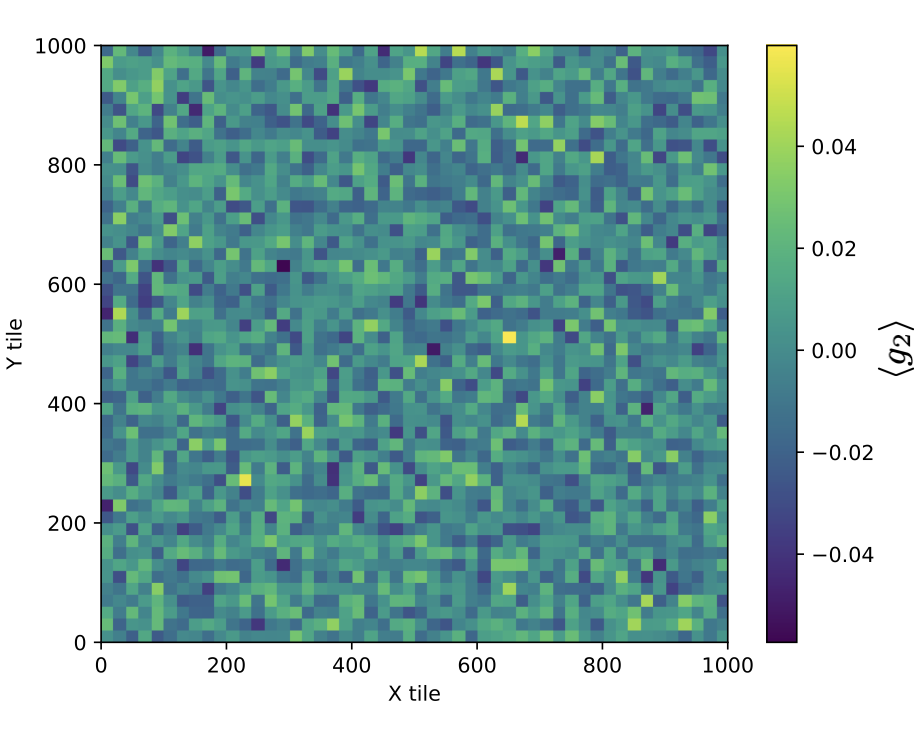}
    \caption{Mean ellipticity components (\emph{left panel:} $g_1; \emph{right:} g_2$) as a function of position on the tile. No complex pattern emerges indicating an absence of additive systematic at the tile level. }
    \label{fig:tile_avg}
\end{figure*}

To gain confidence in the absence of systematics, a common diagnostic is to look at the ellipticity distributions as a function of the core processing units. In our case these are the focal plane, where the image is taken, and the tile, where the exposures are combined for detection. We inspect the average ellipticity binned in focal plane position and tile position in Figs.~\ref{fig:foc_plane_avg} and \ref{fig:tile_avg}. In the case of the focal plane, each galaxy contributes to multiple pixels, appearing at various positions across different exposures. Qualitatively, no pattern seems to emerge in either of the plots. Given the level of PSF leakage, one could have imagined seeing some trends related to the coherent PSF field. 
Having found no global trends visually, we estimate the reduced $\chi^2$ for each histogram as a quantitative check. Assuming pure Poisson noise and testing consistency with zero, the reduced $\chi^2$ values are
\begin{table}[h!]
\centering
\begin{tabular}{lcc}
\hline
 & CCD & Tile \\
\hline
$g_1$ & $20525.7/18000 = 1.14$ & $2565.6/2500 = 1.03$ \\
$g_2$ & $20175.1/18000 = 1.12$ & $2450.4/2500 = 0.98$ \\
\hline
\end{tabular}
\vspace{0.3cm}
\caption{$\chi^2$/d.o.f. values of the mean ellipticities at the CCD and tile levels.}
\end{table}
At the tile level the reduced $\chi^2$ values are close to 1, while at the CCD level they differ slightly. We attribute this to the PSF leakage, as the PSF displays a coherent pattern across the focal plane. Therefore this small excess should be captured by the PSF leakage scheme detailed in \Cref{sec:psf_conta}.

\section{Impact of additive shear biases correlating with galaxy and survey properties}\label{append:correl_bias}

\begin{figure*}[!tb]
    \centering
    \hspace{-2cm}
    \includegraphics[width=0.95\linewidth]{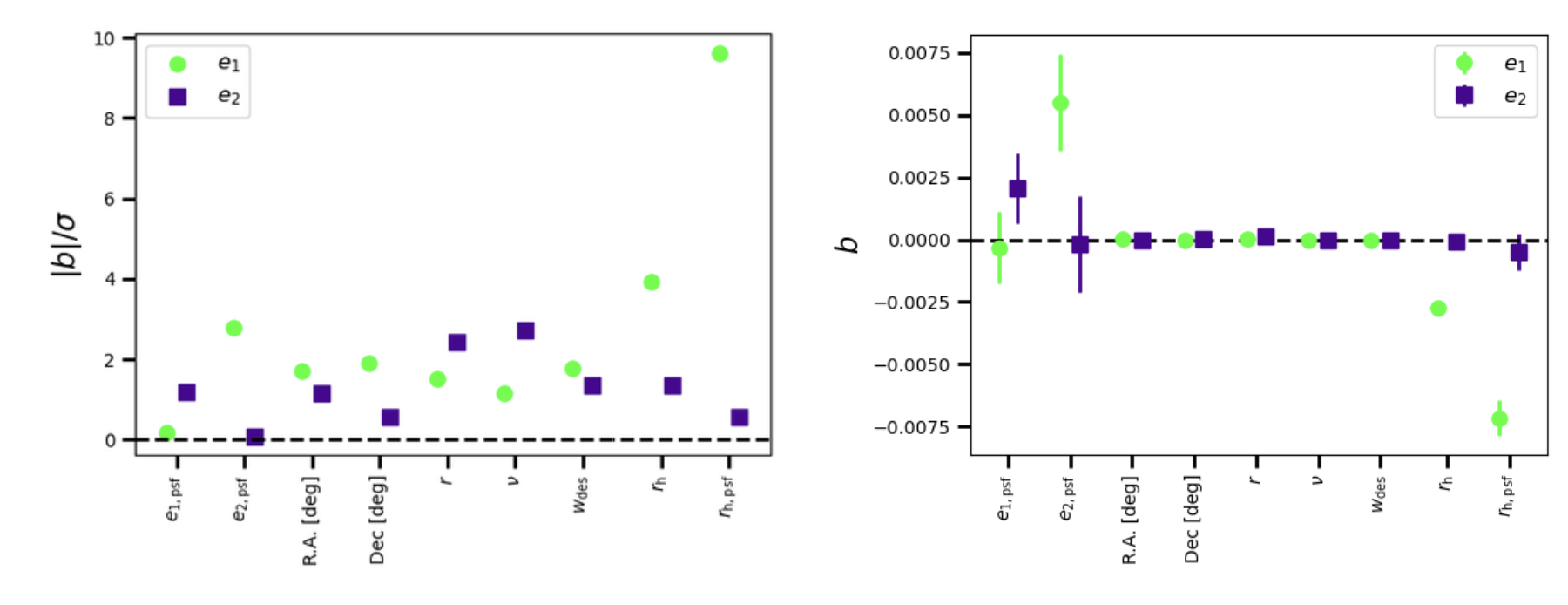}
    \caption{\textit{Left}: Bias value from a linear regression in bins of an observational property $S(x)$. \textit{Right}: Significance of the bias parameter with respect to the fit uncertainty from \cref{eq:ellip_bias}.}
    \label{fig:Clara_m}
\end{figure*}

While the contamination of galaxy ellipticities stemming from a poor removal of the PSF shape is a primary concern due to the PSF induced correlations possibly mimicking the cosmological signal, we can empirically verify if ellipticities correlate with any other observational quantity. To test the robustness of the shear estimate we perform a simple linear regression as a function of observable parameters. In Fig.~\ref{fig:Clara_m} the bias factor $b_i$ is determined through the linear relation
\begin{equation}\label{eq:ellip_bias}
    e^{\mathrm{obs}}_i=\tilde e_{i}+b_i S(x)+c \ ,
\end{equation}
where $e^{\mathrm{obs}}_i$ and $\tilde e_i$ are respectively the observed and true ellipticities.  $S(x)$ is a mean-subtracted galaxy or observational property as a function of position $x$ and $b_i$ is the best fit obtained from the linear-relation $e(S(x))=b_iS(x)+c$. The parameter $b_i$ captures an additive bias trend with respect to a given observable $S(x)$ and should not be
confused with the global multiplicative calibration derived from image simulations in \cref{sec:m_bias}. Its numerical value shown in Fig.~\ref{fig:Clara_m} is also difficult to compare across observables because the dynamical range of $S(x)$ varies from case to case. 
Looking at the significance, the low values of the $(e^\mathrm{obs}_i, \, e_{i}^\textrm{psf})$ correlations arise from the leakage correction applied in \cref{sec:psf_conta}. The $(e^\mathrm{obs}_i, \, e_{j}^{\mathrm{psf}}, \,  [i\neq j])$ correlations have not been calibrated out and are therefore not compatible with 0. Similarly the correlation of $r_{\mathrm{psf}}$ of almost $10 \sigma$ with $e_1$ is an indication  of uncorrected PSF effects at the galaxy level. This dependency is therefore accounted for in the two-point correlation function correction term presented in \cref{seq:galaxy-PSF correlations}. In DES Y3 a correlation of $5.35 \sigma$ with $e_1$ is found for the size ratio, while the significance with $e_2$ is only about $1.5 \sigma$. The similarity of these results is puzzling and further investigations could be relevant to determine if these effects have a common cause. 

The additive bias correlations shown in Fig.~\ref{fig:Clara_m} can be split into three groups. The three PSF quantities are captured by the two-point leakage correction scheme. The RA, DEC and $w_\mathrm{des}$ correlations are not image-level quantities that can generate a well-defined additive shear leakage template; we therefore only report their low bias values here for completeness. 
Finally, for $r$-mag, $\nu$ and $r_\textrm{h}$, we test the auto-correlation due to their contamination to verify that they do not affect the cosmological inference. We expand and formalise the method described in \cite{gattiDarkEnergySurvey2021}. We start by defining the two-point correlation functions of observed ellipticities based on \cref{eq:ellip_bias}, as
\begin{equation}
    \langle e^\mathrm{obs} e^\mathrm{obs}\rangle=\langle \tilde e\tilde e\rangle +2\langle \tilde e \ b\  S(x)\rangle+\langle b S(x) \ b S(x)\rangle \ .
\end{equation}
The cross term can be non-zero when the considered field correlates with shear, such as survey depth for example which locally shifts the redshift distribution or magnification, which can change the galaxy size distribution. A non-zero $b$ is caused by a biased shape measurement process which introduces an additive bias  correlating with survey quantities. In our case, as is done in \cite{gattiDarkEnergySurvey2021}, we are satisfied if the auto-correlation term is less than two percent of the total $\langle e^\mathrm{obs} e^\mathrm{obs}\rangle $ function. We therefore verify that $\langle b S(x)b S(x)\rangle/\langle e^\mathrm{obs}e^\mathrm{obs}\rangle<0.02$ on the scales we use to fit cosmology, giving us an estimation of the contamination due to the auto-correlation of the contamination which is well within the uncertainties.

\section{Catalogue sub-versions}\label{append:cat_versions}

\begin{table*}[!t]
\centering
\caption{\textsc{ShapePipe} catalogue versions and survey properties. $A_\mathrm{obs}$ is the total observed footprint; $A_\mathrm{eff}$ is the effective area after masking (v1.4 variants only).}
\label{tab:v14_survey_properties}
\begin{tabular}{lrrrrrllp{0.25\textwidth}}
\hline\hline
Version & $A_\mathrm{obs}$ & $A_\mathrm{eff}$ & $n_\mathrm{gal}$ & $n_\mathrm{eff}$ & $\sigma_e$ & PSF & Notes & Reference \\
 & (deg$^2$) & (deg$^2$) & ($10^6$) & (arcmin$^{-2}$) & & & & \\
\hline
0.1   & $1{,}500$ & \multicolumn{1}{c}{---} & $40$  & \multicolumn{1}{c}{---} & \multicolumn{1}{c}{---} & PSFex &
  P3 patch & \Axel; \cite{aycoberryUNIONSImpactSystematic2023,robisonShapeDarkMatter2023} \\
1.1   & $3{,}200$ & \multicolumn{1}{c}{---} & $128$ & \multicolumn{1}{c}{---} & \multicolumn{1}{c}{---} & MCCD  & FLAGS=2 & \cite{Hervas_Peters_IA_2024} \\
1.3   & $3{,}200$ & \multicolumn{1}{c}{---} & $84$  & \multicolumn{1}{c}{---} & \multicolumn{1}{c}{---} & MCCD  & + cuts & \cite{liBlackHoleHalo2024,guerriniGalaxyPointSpread2025,zhangPointSpreadFunction2024} \\
\hline
v1.4.5    & $3{,}648$ & $2{,}894$ & $85.0$ & $6.48$ & $0.28$ & PSFex & Initial &  (this work); \cite{milli-vanilli,Daley_B_mode_2026,melody,harmony} \\
v1.4.6.3  & $3{,}648$ & $2{,}894$ & $61.4$ & $4.96$ & $0.27$ & PSFEx & Size cut (fiducial) & \\
v1.4.8    & $3{,}648$ & $2{,}517$ & $53.9$ & $4.98$ & $0.27$ & PSFEx & + star-halo mask & \\
v1.4.11.3 & $3{,}648$ & $2{,}894$ & $78.2$ & $6.26$ & $0.27$ & PSFEx & + FLAGS $\le$ 2 & \\
\hline
\end{tabular}
\tablefoot{$n_\mathrm{eff}$ follows \citet{heymansCFHTLenSCanadaFranceHawaiiTelescope2012}; $\sigma_e$ is the per-component ellipticity dispersion. Both use \texttt{Metacalibration}-weighted galaxy samples. v1.4.11.3 relaxes the \textsc{SExtractor} flag criterion to \texttt{FLAGS} $\le 2$, admitting objects with neighbour-biased photometry (1) or deblending (2). Versions v1.4.5, v1.4.6.3, and v1.4.11.3 share the same footprint (Section~\ref{sec:masking}); v1.4.8 adds stellar-halo masking.}
\end{table*}

In Table \ref{tab:v14_survey_properties} we detail the different sub-versions of the catalogue that were generated, along with their respective cuts and processing choices. In Paper II, some tests on the different versions of the catalogue are made to establish which one has B-mode values compatible with the null hypothesis on the scales of interest. 

\end{appendix}
\end{document}

%% file: macro.tex
\newcommand{\ep}{{e^{\textrm{PSF}}}}                                              
\newcommand{\Tp}{{r_{\rm hlr}^{\textrm{PSF}}}}
\newcommand{\estar}{{e^*}}
\newcommand{\Tstar}{{r_{\rm hlr}^*}}
\newcommand{\esup}[1]{{e^{\textrm{#1}}}}

\newcommand{\size}{\mathcal{R}}
\newcommand{\snr}{\nu_\mathrm{SNR}}
\newcommand{\xisys}[1]{\xi^\mathrm{sys}_{#1}}

\newcommand{\Omegam}{\Omega_{\mathrm{m}}}

%% file: R_g_values.tex
\newcommand{\Rgaa}{0.803}
\newcommand{\Rgbb}{0.801}